\documentclass[a4paper,11pt]{article}
\usepackage{jcappub}
\usepackage{natbib}
\usepackage{aas_macros}
\usepackage{gensymb}
\usepackage{xcolor}

\usepackage{graphics,graphicx, color}
\usepackage{dcolumn, amsmath,amssymb}

\title{Pevatron at the Galactic Center: Multi-Wavelength Signatures from Millisecond Pulsars}

\author[a]{Claire Gu\'epin}
\author[b]{Lucia Rinchiuso}
\author[a]{Kumiko Kotera}
\author[b]{Emmanuel Moulin}
\author[c]{Tanguy Pierog}
\author[a,d,e]{Joseph Silk}
\affiliation[a]{Sorbonne Universit\'es, UPMC Univ. Paris 6 et CNRS, UMR 7095,  Institut d'Astrophysique de Paris, 98 bis bd Arago, 75014 Paris, France}
\affiliation[b]{IRFU, D\'epartement de Physique des Particules, CEA, Universit\'e Paris-Saclay, F-91191 Gif-sur-Yvette, France}

\affiliation[c]{Institut f\"{u}r Kernphysik, Karlsruhe Institute of Technology (KIT), Karlsruhe, Germany }
\affiliation[d]{Department of  Physics \& Astronomy, The Johns Hopkins University, 
3400 N Charles Street, Baltimore, MD 21218, USA}
\affiliation[e]{Beecroft Institute of Particle Astrophysics and Cosmology, Department of Physics,
University of Oxford, Denys Wilkinson Building, 1 Keble Road, Oxford OX1 3RH, UK}

\emailAdd{claire.guepin@iap.fr}
\emailAdd{lucia.rinchiuso@cea.fr}
 
\abstract{
Diffuse TeV emission has been observed by H.E.S.S. in the Galactic Center region, in addition to the GeV gamma rays observed by Fermi. We propose that a population of unresolved millisecond pulsars located around the Galactic Center, suggested as possible candidates for the diffuse Galactic Center excess observed by Fermi, accelerate cosmic rays up to very high energies, and are thus also responsible for the TeV excess. We model analytically the diffusion of these accelerated protons and their interaction with the molecular clouds, producing gamma rays. The spatial and spectral dependences of the gamma rays produced can reproduce the H.E.S.S. observations, for a population of $\sim 10^4-10^5$ millisecond pulsars above the cosmic-ray luminosity $10^{34}\,{\rm erg\,s}^{-1}$, with moderate acceleration efficiency. More precise measurements at the highest energies would allow us  to constrain the properties of the pulsar population, such as the magnetic field or initial spin distributions.
}

\begin{document}

\maketitle

\section{Introduction}

Recent gamma-ray measurements provide  evidence that the Galactic Center (GC) hosts very high energy sources
that produce diffuse gamma-ray emission ranging from GeV to $>10\,$TeV energies \citep{Aharonian06, HESS_GC16, FermiGC17}. Whether this emission results from one single source, tens or even thousands of  sources, or a diffuse population, whether the GeV and TeV observations are connected at all, whether they are produced via similar processes or are the signatures of different particles accelerated in the same sources, whether they stem from leptonic or hadronic models, are all highly debated topics.
Strong arguments have however been put forward in favor of a yet-unresolved population of millisecond pulsars (MSP), being responsible for the GeV gamma rays observed by Fermi known as the Galactic Center excess, through a leptonic channel \citep{FermiGC17}. At higher energies, the H.E.S.S. observations are interpreted as a convincing proof that protons are accelerated up to PeV energies \citep{HESS_GC16}. 

In this work, we connect these salient conclusions in a unified model: we propose that the MSP that are most likely the emitters of the GeV GC gamma rays observed by Fermi are also loaded in baryons, and are thus possible PeV proton accelerators, producing the H.E.S.S. diffuse TeV emission. In this scenario, the pulsars accelerate cosmic-rays up to very high energies. After escaping the sources and diffusing in the Galactic Center region, these accelerated cosmic-rays interact with the molecular clouds during their propagation in the interstellar medium, producing gamma rays. We demonstrate that our model is consistent in terms of energetics and population features. Furthermore, by taking into account spatial diffusion of cosmic rays, we can successfully account for the observations from $100$\,GeV to $>10\,$TeV, and put constraints on key parameters of the millisecond pulsar population. In particular, the cosmic-ray acceleration efficiency within the pulsars, as well as the spatial, magnetic field and initial spin distributions, and the total number of MSP in this population, influence the gamma-ray emission.

We first review in Section~\ref{section:observations} the Fermi-LAT and H.E.S.S. observations, their interpretations available in the literature, and show in Section~\ref{sec:mspulsars} how our millisecond pulsar model can reproduce the derived energetics at first order. Modeling the diffusion of cosmic  rays around the Galactic center and the production of cosmic rays by the MSP are key issues in this study. We examine the diffusion of particles from one source and two populations of MSP in Section~\ref{sec:diff}, study the injection of cosmic-rays by MSP and calculate the associated diffuse gamma-ray flux in Section~\ref{sec:flux}. Section~\ref{sec:discussion} is devoted to a discussion of the results.

\section{Multi-wavelength observations of the Galactic Center}\label{section:observations}

The quality and amount of data towards the GC collected over the last decade from radio to gamma rays have boosted our understanding of high-energy processes taking place in this region (see, for instance, 
Ref.~\citealp{vanEldik15} for a review). The last couple of years have been even more exciting with the measurements in gamma rays of several extended sources, and the refined measurements of GeV-to-TeV diffuse emissions around the Galactic Center. 
We will discuss in this section two major detections that are relevant for the present study: the H.E.S.S. and Fermi-LAT observations of {\it a priori} independent diffuse emissions around the Galactic Center, and the corresponding interpretations that are being discussed in the literature. We caution that the observations that are relevant to us exclude the GeV and TeV sources, 1FGL J1745-290 and HESS J1745-290, respectively, which are spatially coincident with the supermassive black hole Sagittarius A*. This object is not considered to be the source of the GeV-TeV diffuse emission that  we aim to model. 
We first recall some basics of the structure of the Galactic Center and of the millisecond pulsar distribution that are relevant to understanding the interpretations of the high-energy gamma-ray emission.

\subsection{ The Galactic Center region} \label{section:GCregions}
\begin{figure}[t]
\centering
\includegraphics[width=0.6\textwidth]{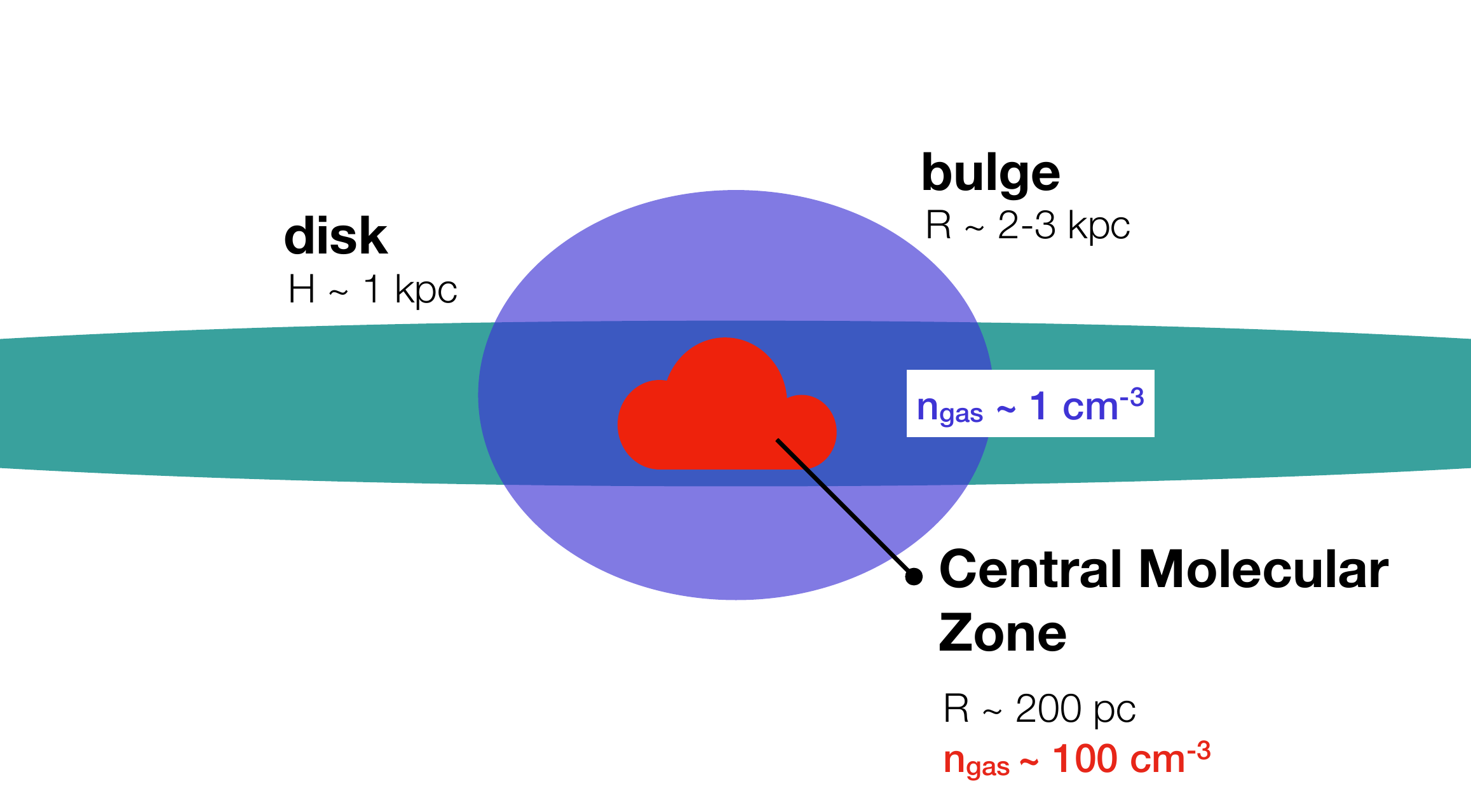}
\caption{Sketch of the regions of the Galactic Center at play in our model, with indications on the approximate size, gas density and millisecond pulsar numbers, as detailed in Section~\ref{section:GCregions}.}\label{Fig:GCsketch}
\end{figure}

Radio observations of pulsars combined with information from star formation rates show that the bulk of the pulsar population is concentrated in the Galactic disk, and that it could contain thousands of objects \citep{Levin13,Lorimer04,Lorimer13}. The Galactic disk can be modeled as a cylinder of height $\sim 1\,$kpc, and of gas density $n_{\rm gas}\sim 1\,{\rm cm}^{-3}$ (see Fig.~\ref{Fig:GCsketch}). 

The inner few kiloparsecs of our Galaxy are commonly referred to as the {\it bulge} of the Galaxy. It consists of an elongated structure stretched over $2-3\,$kpc, populated by old ($\sim 10\,$Gyr old) stars, and thus putatively hosting an important population of millisecond recycled pulsars \citep{Zoccali16,FermiGC17}. Except for the very inner region 
hosting molecular clouds known as the Central Molecular Zone (CMZ), the gas density in the bulge can be roughly approximated to be similar to that in the disk.

One specificity of the Galactic Center region  is that it is filled with giant molecular clouds, that represent about 10\% of the total gas amount of the Galaxy (see Ref.\cite{Mills17} for a review). The emission in this region is dominated by non-thermal  radiation from accelerated particles, with several identified powerful objects such as supernova remnants and pulsar wind nebulae. As a consequence, the energy density in the CMZ is estimated to be an order of magnitude larger than that of the average Galactic cosmic rays. The molecular clouds, with a mean gas density $n_{\rm gas}\sim 100\,{\rm cm}^{-3}$, are believed to be prime targets for the production of the observed gamma rays. The radio, infrared and submillimeter images reveal a ridge-like, elongated morphology for the gas distribution, mostly concentrated in a radius of $\lesssim 200\,$pc around the Galactic Center. 

\subsection{The diffuse TeV emission}\label{section:TeVobs}
Deep observations of the GC region carried out  by the H.E.S.S. collaboration 
 revealed an extended diffuse emission over a few hundred parsecs around the Galactic Center from $-1.1^\circ$ to $+1.5^\circ$ in Galactic longitude~\cite{Aharonian06}.  The statistics accumulated over 10 years together with improved analysis techniques have enabled us to map this region with increased accuracy, and have revealed  diffuse emission in the inner 50 pc around Sagittarius A*, reaching gamma-ray energies $E>10\,$TeV \cite{HESS_GC16,Abdalla:2017xja}. This region, hereafter referred to as the inner 50 pc region, is defined as an annulus centered of Sgr A$^\star$ of inner and outer radii of 0.15$^{\circ}$ and 0.45$^{\circ}$, respectively. Angles between 304$^{\circ}$ and 10$^{\circ}$ in Galactic coordinates are excluded from the integration region.
This emission is spatially correlated with the CMZ, and hence points towards the acceleration of protons in this region. Indeed, a leptonic scenario with electrons and positrons that undergo Inverse Compton scattering  off the radiation field is unlikely, as the leptons would dominantly suffer severe synchrotron radiative losses that would prevent them from propagating over the scale of the CMZ. A hadronic scenario seems more favorable in this perspective,  where energetic protons interacting with the gas in the interstellar medium produce very-high-energy (VHE, E $\gtrsim$ 100 GeV) gamma rays from $\pi^0\rightarrow\gamma\gamma$ decay. The total $\gamma$-ray luminosity injected in this region is measured to be of order of $L_{\gamma>1\,{\rm TeV}} \sim 5 \times 10^{34}\,{\rm erg\,s}^{-1}$.

The detection of VHE gamma rays in the 10 TeV energy range requires the acceleration of CR protons to PeV energies, which implies either one or a population of accelerators of such particles, called as {\it pevatrons}, in the Galactic Center region. The central supermassive black hole Sagittarius A* 
could accelerate ultra-relativistic protons to PeV energies, thus acting as a Pevatron. In the scenario of a central single PeV source, the radial dependency of the CR proton profile up to a few hundred parsecs from Sagittarius A* suggests continuous injection of protons over timescales of at least thousands of years. It was initially suggested that a single supernova explosion could explain this emission, by the injection and diffusion of particles, and their interaction with the molecular clouds. However, a single supernova can hardly sustain efficient PeV proton acceleration over such a timescale~\cite{Bell:2013kq}.

\subsection{The diffuse GeV emission}\label{section:GeVobs}
A high-energy gamma-ray excess with respect to the interstellar emission models has been detected using Fermi-LAT observations  with a spatial extension up to about 20 degrees from the Galactic Center (see, for instance, Refs.\citep{Goodenough09, Abazajian12, Hooper13, Abazajian14, Calore15, Daylan16}). Several gamma-ray emission scenarii have been suggested, however there is no definite conclusion on the origin of the excess.  Among them are dark matter annihilations in the inner region of the Galactic dark matter halo \citep{Goodenough09,Hooper13,Abazajian14,Calore15,Daylan16}, as well as outflows from the supermassive black hole Sagittarius~A$^\star$ injecting energetic cosmic-ray protons \citep{Carlson14} or leptons \citep{Petrovic14,Gaggero15} in the interstellar medium from outflows. While the former may be in tension with the non-observation of gamma-ray excesses towards dwarf galaxy satellites of the Milky Way \citep{PhysRevLett115231301}, the latter would hardly reproduce the morphology of the Galactic Center excess. An alternative hypothesis is the presence of an additional SNR population that could steadily inject protons \citep{Gaggero15,Carlson16}, being however not observed at any other wavelength so far. More recently, a hypothetical population of MSP in the Galactic bulge has been shown to well match the morphology of the Galactic Center excess \cite{Wang05,Cholis15,Lee15,Bartels16}. The presence of this unresolved pulsar population has been independently put forward by Fermi-LAT \citep{FermiGC17} using 7.5 years of data of Pass 8 analysis\footnote{The unresolved bulge pulsar population is robustly detected against the underlying interstellar emission models possibly including the Fermi bubble component~\citep{FermiGC17}.}.

Besides the pulsar population of the Galactic disk, an additional distinct bulge pulsar population is needed, for radial distance $r<3$\,kpc from the Galactic Center.
The disk population follows a Lorimer Galactocentric spatial distribution $\rho(R)\approx R^n e^{-(R/\sigma)}$ with $n=2.35$ and $\sigma=1.528$ kpc, and a distribution as a function of the distance from the Galactic disc $\rho(z)\approx e^{-(|z|/z_0)}$ with scale height $z_0=0.70$ kpc.
The luminosity function for the gamma-ray emission is modelled as a power-law with slope $-1.7$ in the luminosity range $[10^{33},10^{36}]\,$erg\,s$^{-1}$. The number of expected pulsars in the disk was derived to be $N_{\rm d}=[4000-16000]$, based on the known pulsars and the unassociated 3FGL sources compatible with pulsar characteristics. The additional distinct bulge pulsar population is well described by a spherically symmetric distribution ${\rm d}N/{\rm d}r\propto r^{-\alpha_{\rm b}}$, with $\alpha_{\rm b}=2.6$.
The bulge luminosity function is modelled as for the disk and the normalization is determined in order to reproduce the Galactic Center excess. The number of pulsars in the bulge is estimated to be in the range $N_{\rm b}=[800-3600]$ in the luminosity range $[10^{33},10^{36}]\,$erg s$^{-1}$.
Such an estimate can be affected by systematic uncertainties in the modelling of the MSP populations. Among them are the construction of the interstellar emission model, the modelling of the MSP disk population, and the assumed luminosity functions of the disk and bulge populations. Interestingly, in a recent study \cite{Ploeg17}
the authors derived $N_{\rm b}$ = (4.0$\pm$0.9)$\times$10$^4$ for MSP luminosities greater than $10^{32}\,$erg\,s$^{-1}$. Extrapolating the derived luminosity function from Ref.~\cite{FermiGC17}  down to $10^{32}\,$erg \,s$^{-1}$  provides compatible with the results of Ref.~\cite{Ploeg17} within errors.

A recent bayesian study of gamma-ray emitting MSP \cite{Bartels18} suggest the presence of $2\times 10^4-10^5$ MSP in the Galactic disk, a number that is in agreement with the population derived from radio catalogs \cite{Levin13}. The authors find that the luminosity function in the disk population preferably follows a Lorimer power-law profile as we assume in the current work. They report that they lack sensitivity to place strong constraints on the bulge population of MSP.

\section{Millisecond pulsars as pevatrons}\label{sec:mspulsars}
The evidence of PeV protons in the Galactic Center, together with the report that a millisecond pulsar population may be responsible of the Galactic Center diffuse emission observed by Fermi-LAT, led us to elaborate the following scenario. A millisecond pulsar population emits the diffuse Fermi GeV gamma rays via leptonic processes, and the diffuse TeV excess observed by H.E.S.S. via hadronic processes, hence acting as pevatrons. In this scenario, MSP accelerate protons up to very high energies, that can reach PeV energies for initial spin periods of $P_{\rm i}\sim 1\,{\rm ms}$ and dipole magnetic fields $B\gtrsim 10^9\,{\rm G}$. These cosmic-rays interact with the interstellar medium and molecular clouds through hadronic processes and produce neutral pions that decay into gamma rays. The millisecond pulsar population is characterized by a spatial distribution around the Galactic Center, and by period, magnetic field and age distributions.

The diffusion of cosmic rays emitted from each pulsar leads to a typical radial extension of the cosmic-ray density that can be compared to the data. We model the propagation of protons in the turbulent Galactic magnetic field by following the estimates of Ref.~\cite{Blasi12a} for the diffusion coefficient, as we will explain in detail in Section~\ref{sec:CR_dens}. Typical estimates of proton-proton interaction and diffusion timescales, written as
\begin{eqnarray}
t_{pp} &=& 1/c n_{\rm H} \sigma_{pp} \, , \\
& \sim & 10^{13} \, {\rm s} \,\left({n_{\rm H}}/{100\,{\rm cm}^{-3}}\right)^{-1} \, , \nonumber
\end{eqnarray}
and
\begin{eqnarray}
t_{\rm diff} &=& r_{\rm diff}^2/2D \, , \\
& \sim & 10^{11} \, {\rm s} \, \left({r_{\rm diff}}/{200\,{\rm pc}}\right)^{2}  \,,\nonumber
\end{eqnarray}
respectively, where $\sigma_{pp} \simeq 50\,{\rm mb}$ is the hadronic cross section for a proton energy of $E = 10^{14}\,{\rm eV}$, $n_{\rm H}$ the gas density (see Section~\ref{section:GCregions}) and $D \simeq 10^{30} \,{\rm cm}^2\,{\rm s}^ {-1}\,E_{14}$ is the diffusion coefficient for protons at $E = 10^{14}\,{\rm eV}$ (see Section~\ref{sec:CR_dens} for more details).
 As $t_{pp} > t_{\rm diff}$, one expects a large radial extension of the cosmic-ray density distribution.

In this study, we only consider the impact of proton-proton interactions and neglect other energy loss processes, as synchrotron or inverse Compton processes. The typical interaction timescales of these processes are respectively $t_{\rm syn}^{-1} \sim 4/3 \,\sigma_{{\rm T},p} c \gamma_p^2 U_B E^{-1}$ and $t_{\rm IC}^{-1} \sim 4/3 \,\sigma_{{\rm T},p} c \gamma_p^2 U_{\rm rad} E^{-1}$ (in the Thomson regime), where $\sigma_{{\rm T},p}$ is the Thomson cross section for protons, $U_{\rm rad}$ is the CMB energy density and $U_B = B^2/8\pi$ is the magnetic energy density. We obtain the estimates $t_{\rm syn} \sim 7 \times 10^{14}\,{\rm yr} \, E_{13}^{-1} B_{-4}^{-2}$ and $t_{\rm IC} \sim 1 \times 10^{18}\,{\rm yr} \, E_{13}^{-1}$ for $E_{13} = 10^{13}\,{\rm eV}$, $B_{-4} = 100\,\mu{\rm G}$ and $U_{{\rm rad}} \sim 0.3 \,{\rm eV\,cm}^{-3}$. These are well above the typical proton-proton energy-loss timescale $t_{pp} \sim 5 \times 10^7\,{\rm yr}\,n_{{\rm H},1}$ for $E_{13} = 10^{13}\,{\rm eV}$ and $n_{{\rm H},1} = 1\,{\rm cm}^{-3}$, which confirms that these processes are sub-dominant when compared to proton-proton interactions.

Considering the MSP population inferred to explain the diffuse GeV emission, we assess if the energy reservoir in this population is sufficient to reach the level required to fit the gamma-ray flux. In the following, $B$ is the dipole magnetic field strength of the star, $R_\star$ its radius and $P_{\rm i}$ the initial spin period. We consider that the electromagnetic luminosity of pulsars \citep{ST86,Arons03}
\begin{eqnarray}
|\dot{E}_{\rm rot}| &=& 16\pi^4 B^2 R_{\star}^6 P^{-4}/9c^3 \, ,\\
&\simeq& 6.4 \times 10^{36}\,{\rm erg\,s}^{-1}\,B_9^2 R_{\star,6}^6 P_{{\rm i},-3}^{-4} \, , \nonumber
\end{eqnarray}
is converted to kinetic luminosity $\dot{N} E$, with efficiency $\eta_{\rm acc} \le 1$. The particle rest mass power is $\dot N mc^2 \,\equiv\,\dot N_{\rm GJ} ( 2\kappa\,m_e + A m_{\rm p}/Z) c^2$, where 
\begin{eqnarray}
\dot N_{\rm GJ} &\sim& \mathcal{A}_{\rm PC}\, \rho_{\rm GJ} \,c/e  = 2 \pi^2 B R_{\star}^3 P^{-2} / ec \, ,\\
&\simeq& 1.4 \times 10^{33}\,{\rm s}^{-1}\,B_9 R_{\star,6}^3 P_{{\rm i},-3}^{-2} \, ,\nonumber
\end{eqnarray}
is the Goldreich-Julian rate \citep{Goldreich69,Arons03}, with $\mathcal{A}_{\rm PC} \simeq 2 \pi^2 R_\star P^{-1}/c $ the area of a polar cap and $ \rho_{\rm GJ} \simeq B P^{-1} / c$ the Goldreich-Julian charge density \citep{Goldreich69}. Therefore millisecond-pulsars can accelerate protons up to very high energies \citep{Kotera15}: 
\begin{eqnarray}\label{eq:E0}
E_0 &=& \eta_{\rm acc} |\dot{E}_{\rm rot}| / \dot{N} \, , \nonumber\\
&\sim& 1.4\times10^{15}\,{\rm eV}\, \eta_{\rm acc}\,\kappa_3^{-1}(1+m_p/2m_e\kappa_3)^{-1} B_{9}R_{\star,6}^3P_{{\rm i},-3}^{-2} \, .
\end{eqnarray}
where $\kappa$ is the pair multiplicity, which can range between $10-10^8$ in theory (a highly debated quantity) and $\eta_{\rm acc}=1$. For $\kappa \sim 10^3$, most of the pulsar power goes into ions, as $m_p/2\kappa m_e \sim 0.9$. Taking into account the pulsar spin-down, characterized by the spin-down timescale $t_{\rm sd} = 9Ic^3P^2 / 8\pi^2B^2R_\star^6 \sim 9.8 \times 10^{7}\,{\rm yr}\,I_{45}B_{9}^{-2}R_{\star,6}^{-6}P_{{\rm i},-3}^{2}$,  the cosmic-ray energy at a time $t$ is $E_{\rm CR}(t) = E_0 (1+t/t_{\rm sd})^{-1}$. Following Refs.~\citep{Blasi00,Arons03}, the cosmic-ray luminosity in protons is given by
\begin{eqnarray}\label{eq:LCR}
L_{\rm CR}(t) &=& \frac{9}{4} \frac{c^2 I}{eBR_\star^3} E_{\rm CR}(t) (t+t_{\rm sd})^{-1} \, , \\
&\simeq& 3.1 \times 10^{36}\,{\rm erg\,s}^{-1}\, \eta_{\rm acc}\, \kappa_3^{-1}(1+m_p/2m_e\kappa_3)^{-1} B_{9}^{2} \, R_{\star,6}^{6} \, P_{{\rm i},-3}^{-4} \,(1+t/t_{\rm sd})^{-2} \nonumber \, ,
\end{eqnarray} 
where the latter value is obtained for  $\eta_{\rm acc} = 1$ and $ \kappa = 10^3 $.

In the following we neglect the potential interaction of accelerated cosmic rays in the vicinity of the source, with the ambient photon fields or hadronic debris, which is out of the scope of the present study. From the millisecond pulsar luminosity $L_{\rm MSP} \sim 10^{36}\,{\rm erg\,s}^{-1}$, and the luminosity in baryons $L_p = \eta_p L_{\rm MSP}$, where $\eta_p$ is the fraction of the pulsar luminosity channelled into protons, we have $L_{{\rm MSP,tot}} = N_{{\rm MSP}} L_{\rm MSP} $ where $N_{{\rm MSP}}$ is the number of MSP is the region considered. Therefore the gamma-ray luminosity $L_{\gamma}$ related to proton-proton interactions is
\begin{eqnarray}
L_{\gamma} &\sim& \tau_{pp}\eta_p  L_{{\rm MSP,tot}} \,, \\
&\sim&  10^{36}\,{\rm erg\,s}^{-1} \,\eta_p \left(\frac{N_{\rm MSP}}{100}\right) \left(\frac{r_{{\rm diff}}}{200\,{\rm pc}}\right)^2 \left(\frac{n_{{\rm H}}}{100\,{\rm cm^{-3}}}\right)\,, \nonumber
\end{eqnarray}
where $\tau_{pp} = t_{\rm diff}/t_{pp}$. Note that the diffuse excess observed by H.E.S.S. is about $L_{\gamma>1\,{\rm TeV}} \sim 5 \times 10^{34}\,{\rm erg\,s}^{-1}$ in the inner 50 pc region, thus the energetic budget estimated above is sufficient to explain the diffuse excess, and leaves room for low injection rate and inefficient sources.

Considering this population of MSPs, we predict the gamma-ray flux profile as a function of distance from the Galactic Center and the inferred cosmic-ray density, as well as the TeV gamma-ray flux energy spectrum in the inner 50 pc region. In order to fit these predictions to the observational data, only a limited number of free parameters are required to be determined: namely, the magnetic field distribution $F_B(B)$, the acceleration efficiency $\eta_{\rm acc}$ and the number of MSP in the population considered.

\section{Cosmic-ray spatial density distribution}\label{sec:diff}
The diffusion of cosmic rays is the key process to estimate their density and its spatial dependency. First, we consider the case of one source and generalize our results to the case of two different MSP populations, in the Galactic bulge and in the disk, respectively. In the following, we focus on the case of accelerated protons.

\subsection{Cosmic-ray density for a single source}\label{sec:CR_dens}
After escaping from a source, cosmic rays diffuse and interact with the surrounding medium. Following \cite{Blasi12a}, we can model the diffusive propagation of protons with the diffusive transport equation
\begin{equation}\label{eq:transport}
\frac{\partial n(E,\vec r,t)}{\partial t}=\nabla\left[D(E)\nabla n(E,\vec r,t)\right] - \Gamma^{\rm sp}(E) n(E,\vec r,t) + N(E) \delta(t-t_{s})\delta^{3}(\vec r - \vec r_{s}),
\end{equation}
where cosmic rays are injected at a time $t_{s}$ from a point source located at $\vec r_{s} = (x_{s},y_{s},z_{s})$, with a spectrum $N(E)$; $n(E,\vec r,t)$ is the density of particles with energy $E$ at the location $\vec r$ and time $t$, $D(E)$ is the diffusion coefficient assumed to be spatially constant and $\Gamma^{\rm sp}(E) $ is the spallation rate of protons. As explained in Section~\ref{sec:discussion}, we neglect proton energy losses, which are typically described by the term $\partial  \left[ P(E) n(E,\vec r,t) \right] /\partial E$. The energy-dependent diffusion coefficient writes 
\begin{equation}\label{eq:diff}
D(E) = 10^{28} D_{28} \left( \frac{R}{3\,{\rm GV}}\right)^{\delta} \rm cm^{2}\, s^{-1} \, ,
\end{equation}
where $R=E/Z$ is the rigidity (with $E$ in eV). The best fit to the existing data of boron-to-carbon ratio is obtained for $D_{28}/H_{\rm kpc}= 1.33$ with $\delta=1/3$ (Kolmogorov-type) \cite{Kolmogorov41, Aguilar16, GALPROP}; $H_{\rm kpc}=3\,{\rm kpc}$ is the halo height in kpc.

The rate of spallation $\Gamma^{\rm sp}(E) $ depends on the gas density $n_{\rm gas}$, the nucleus velocity $v$ (we can assume $v=c$) and the cross section $\sigma_{{\rm pp}}$
\begin{equation}\label{eq:spallation}
\Gamma^{\rm sp}(E) = n_{\rm gas} \,c\, \sigma_{{\rm pp}} \,.
\end{equation}
 At GeV energies and above, the spallation cross-section can be well parametrized by $\sigma_{\rm pp}(E) \simeq 30 \{0.95 + 0.06 \ln[ (E-m_p c^2)/1\,{\rm GeV}]\} \,{\rm mb}$ \citep{Aharonian04}.

\begin{figure}[t]
\centering
\includegraphics[width=0.49\textwidth]{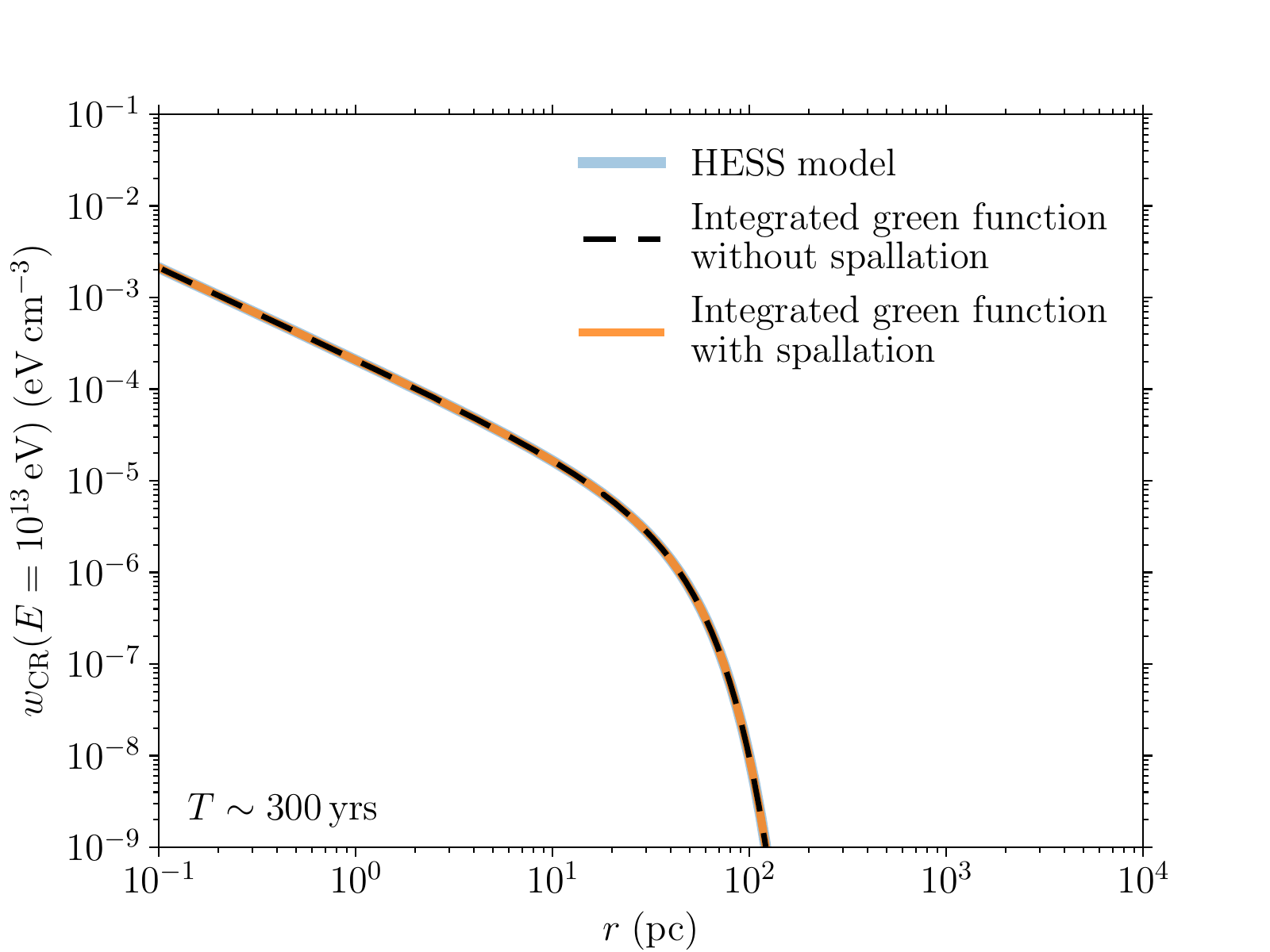}
\includegraphics[width=0.49\textwidth]{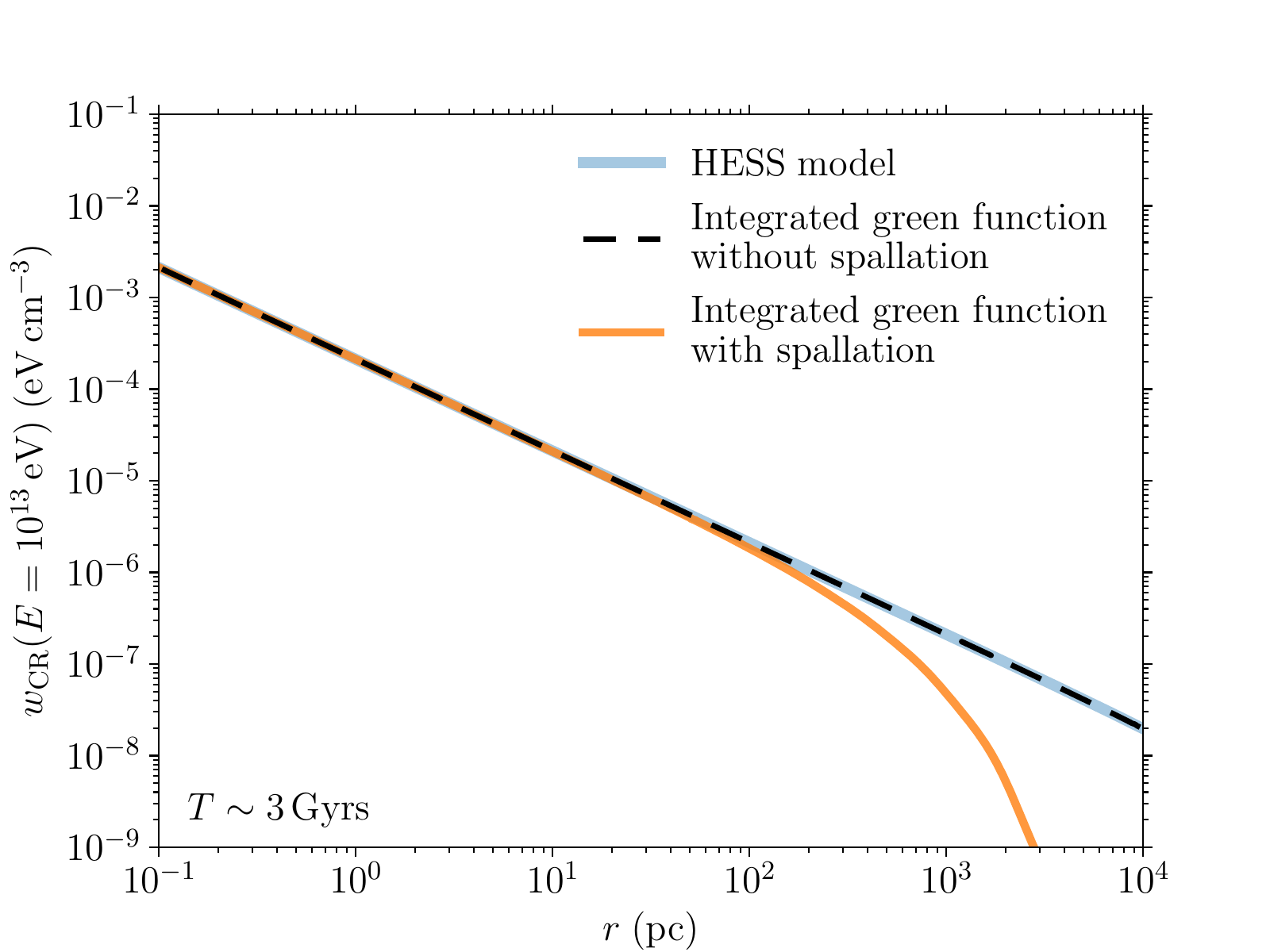}
\caption{Cosmic-ray density for one source, from \cite{HESS_GC16} (blue), obtained with the integrated Green function without spallation (dashed black), or with the integrated Green function with spallation (orange), for $E=10^{13}\,{\rm eV}$. We compare the results for continuous injection times $T\sim 300\,$yrs ({\it left}) and $T\sim 3\,$Gyrs ({\it right}). The three formalisms agree out to distances of a few 100 pc.}\label{Fig:CR_Dens_onesource}
\end{figure}

The following Greens function is a solution of Eq.~(\ref{eq:transport}) without boundary conditions
\begin{eqnarray}\label{eq:green}
{\cal G}(\vec r, t; \vec r_{s},t_{s}) &=& \frac{N(E)}{\left[ 4 \pi D(E) \tau \right]^{3/2}} \exp\left[ -\Gamma^{\rm sp}(E) \tau \right]
\exp\left[ -\frac{|\vec{r}-\vec{r_s}|^{2}}{4 D(E) \tau}\right] \,,
\end{eqnarray}
where $\tau = t-t_{s}$ \citep{Blasi12a}. For a constant source injection rate $\dot{Q}_p(E)$ during the time $T$, we can calculate the cosmic-ray density, at a time $t=T$ and position $\vec r$, by integrating over the injection time $t_{\rm inj}$
\begin{eqnarray}
w_{\rm CR}(E,\vec r,t) &=& \int_{t_{\rm inj}=0}^{t} {\rm d}t_{\rm inj} \dot{Q}_p(E) \,{\cal G}(\vec r, t; \vec r_{s},t_{\rm inj}) \, , \\
&=& \int_{t'=0}^{t} {\rm d}t' \dot{Q}_p(E) \,{\cal G}(\vec r, t'; \vec r_{s},0) \, .\nonumber
\end{eqnarray}
Assuming spherical symmetry and continuous injection over timescale $T =t \ge t_{\rm diff}$, the cosmic-ray density for one central source for radial distances $r <r_{\rm diff}$ writes~\cite{HESS_GC16}
\begin{equation}\label{eq:CRdens_HESS}
w_{\rm CR}(E, r,t) = \frac{\dot{Q}_p(E)}{4\pi D(E)r}\,{\rm erfc}\left(\frac{r}{\sqrt{4D(E)t}}\right) \,.
\end{equation} 
The diffusion radius $r_{\rm diff}$ is assumed to be of $\sim$ 200 pc following the spatial extension of the TeV emission measured by H.E.S.S.~\cite{HESS_GC16}. The corresponding radial cosmic-ray densities obtained in the above-mentioned cases are plotted in Fig.~\ref{Fig:CR_Dens_onesource}, with and without the spallation process. The solution used in Ref.~\cite{HESS_GC16} is accurate over a large range of distances. 

At large distances from the location of the source ($r \gg 100\,{\rm pc}$), we see the effect of spallation for times longer than the spallation interaction timescale $t_{\rm pp} \sim 10^{13}\,{\rm s} \sim 0.2\,$Myrs (for $n_{\rm p} = 100 \,{\rm cm}^{-3}$ and $E = 10^{13}\,{\rm eV}$).
Therefore this cosmic-ray density integrated over injection time is compatible with a solution $\propto r^{-1}$ close to the central source, typically at distances smaller than $100\,{\rm pc}$. As expected, the general solution of the diffusion equation derived in the cylindrical case converges towards the spherical-case solution for time-scales lower than several thousand years.

\subsection{Millisecond-pulsar distributions}
We consider two distinct populations of millisecond-pulsars, one in the bulge and one in the disk, and use the spatial distributions derived in Ref.~\cite{FermiGC17}. 

\paragraph{In the bulge:} the distribution of MSPs is described by $F_{\rm b}(r_s,\theta,\phi) = K_{\rm b} r^{-\alpha_{\rm b}}$ in spherical coordinates, with $K_{\rm b}$ a normalization constant and $\alpha_{\rm b} = 2.6$. By normalizing this distribution to the total number of millisecond-pulsars in the bulge $N_{\rm b}$, we obtain 
\begin{equation}
F_{\rm b}(r_s,\theta,\phi) = \frac{(3-\alpha_{\rm b}) N_{\rm b}}{4 \pi r_{\rm max}^{3-\alpha_{\rm b}}} \,r_s^{-\alpha_{\rm b}} \quad {\rm for} \; 0 < r_s < r_{\rm max}  \, ,
\end{equation}
where $r_{\rm max} = 3.1 \times 10^3 \, {\rm pc} $ is the radial extension of the bulge~\cite{1996A&ARv7289M}. Above $r_{\rm max}$, the disk contribution dominates over the bulge one. In this region, the precise behaviour of the radial dependency of the bulge distribution is neglected. The radial distribution normalized to $1$ is therefore $F_{\rm b}(r_s) = (3-\alpha_{\rm b}) \,r_s^{2-\alpha_{\rm b}} / r_{\rm max}^{3-\alpha_{\rm b}} $ for $0 < r_s < r_{\rm max}$. 

\begin{figure}[t]
\centering
\includegraphics[width=0.5\textwidth]{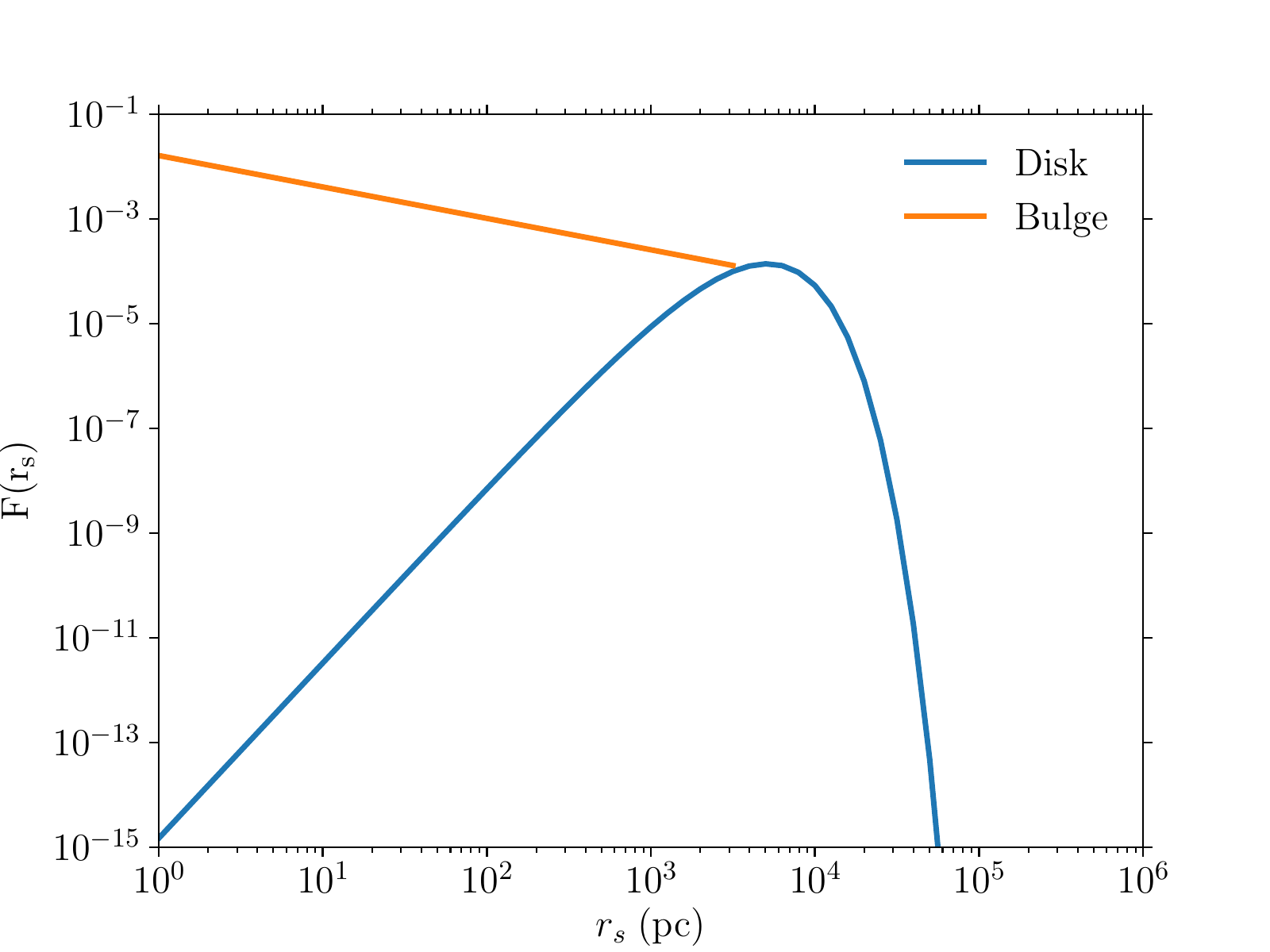}
\caption{ Normalized radial distribution functions of the bulge (orange line) and disk (blue line) populations of MSP.}\label{Fig:Pulsar_Dist}
\end{figure}

\paragraph{In the disk:} the distribution of MSPs, normalized to the total number of millisecond-pulsars in the disk $N_{\rm d}$, is described by 
\begin{equation}
F_{\rm d}(r_s,\theta,z) = \frac{ r_s^n \exp(-r_s/\sigma) \exp(-|z_s|/z_0)\,N_{\rm d}}{4 \pi z_0 \sigma^{n+2} \Gamma(n+2)} \, ,
\end{equation}
in cylindrical coordinates, with $n = 2.35$, $\sigma = 1.528 \times 10^3 \, {\rm pc}$ and $z_0 = 700 \, {\rm pc}$. The radial distribution normalized to $1$ is therefore $F_{\rm d}(r_s) = r_s^{n+1} \exp(-r_s/\sigma) / \sigma^{n+2} \, \Gamma(n+2)$. The two radial distributions normalized to $1$ are illustrated in Fig.~\ref{Fig:Pulsar_Dist}.

\subsection{Total cosmic-ray density}\label{Sec:Total_CR_dens}
First, we focus on the impact of the spatial distribution of MSP on the cosmic-ray density profile, and thus consider a continuous injection of cosmic-rays from each pulsar, during $T \sim 3\,{\rm Gyrs}$, and an observation time $t=T$. This preliminary assumption of continuous injection, which is not realistic in the case of MSP, should be considered as a preliminary step required to study the cosmic-rays radial distribution. Therefore, we assume that the cosmic-ray density for one source is well described by Eq.~(\ref{eq:CRdens_HESS}), where we neglect the error function component. As shown in Section~\ref{sec:CR_dens}, this approximation is reasonable for a continuous cosmic-ray injection from the source, and for short distances from the central source (see Fig.~\ref{Fig:CR_Dens_onesource}). The total cosmic-ray density is calculated analytically by integrating the one-source density over the distribution of millisecond-pulsars in the bulge and the disk. 

\paragraph{In the bulge:} the total cosmic-ray density is given by
\begin{eqnarray}
w_{\rm CR,tot}(E,r,t) &=& \int_{r_s = 0}^\infty \int_{\theta = 0}^\pi \int_{\phi = 0}^{2\pi} r_s^2 {\rm d}r_s \sin\theta {\rm d}\theta {\rm d}\phi  \, F(r_s,\theta,\phi) \, w_{\rm CR}(E,|\vec{r}-\vec{r}_s|,t) \, , \nonumber \\
&=& \frac{(3-\alpha_{\rm b}) \,\dot{Q}_p(E)\,N_{\rm b}}{16\pi^2 D(E) \, r_{\rm max}^{3-\alpha_{\rm b}}} \int_{r_s = 0}^{r_{\rm max}} \int_{\theta = 0}^\pi \int_{\phi = 0}^{2\pi} \frac{r_s^{2-\alpha_{\rm b}} {\rm d}r_s \sin \theta {\rm d}\theta {\rm d}\phi}{\sqrt{r^2+r_s^2-2rr_s\cos\theta}} \, .
\end{eqnarray}
For $r<r_{\rm max}$
\begin{eqnarray}\label{Eq:wCR_int_dist}
w_{\rm CR,tot}(E,r,t) &=& \frac{\dot{Q}_p(E)(3-\alpha_{\rm b})\,N_{\rm b}}{4\pi D(E)(2-\alpha_{\rm b}) \, r_{\rm max}} \left[ 1 - \frac{1}{3-\alpha_{\rm b}} \left(\frac{r}{r_{\rm max}}\right)^{2-\alpha_{\rm b}} \right]\, ,
\end{eqnarray}
and for $r \geq r_{\rm max}$
\begin{eqnarray}
w_{\rm CR,tot}(E,r,t) &=& \frac{ \dot{Q}_p(E)\,N_{\rm b}}{4 \pi D(E) r} \, .
\end{eqnarray}

\paragraph{In the disk:} the total cosmic-ray density is given by
\begin{eqnarray}
w_{\rm CR,tot}(E,r,t) &=& \int_{r_s = 0}^\infty \int_{\theta = 0}^{2 \pi} \int_{z_s = -\infty}^{\infty} r_s {\rm d}r_s {\rm d}\theta {\rm d}z_s  \, F(r_s,\theta,z_s) \, w_{\rm CR}(E,|\vec{r}-\vec{r}_s|,t) \, , \nonumber \\
&=& \frac{\dot{Q}_p(E)\,N_{\rm d}}{16\pi^2 D(E) z_0 \sigma^{n+2} \Gamma(n+2)}  \nonumber \\ 
&& \times \int_{r_s = 0}^\infty \int_{\theta = 0}^{2 \pi} \int_{z_s = -\infty}^{\infty} \frac{r_s^{n+1} {\rm d}r_s {\rm d}\theta {\rm d}z_s \exp(-r_s/\sigma) \exp(-|z_s|/z_0) }{\sqrt{r^2+r_s^2-2 r r_s \cos \theta + (z-z_s)^2}} \, .
\end{eqnarray}
Integrating over $\theta$ we obtain
\begin{eqnarray}
w_{\rm CR,tot}(E,r,t)
&=& \frac{\dot{Q}_p(E)\,N_{\rm d}}{\pi^2 D(E) z_0 \sigma^{n+2} \Gamma(n+2)} \int_{r_s = 0}^\infty \int_{z_s = 0}^{\infty} {\rm d}r_s  {\rm d}z_s \nonumber \\ 
&& \times \frac{ r_s^{n+1}  \exp(-r_s/\sigma) \exp(-|z_s|/z_0)}{(r-r_s)^2+z_s^2} \,\mathcal{K}\left(\frac{-4rr_s}{(r-r_s)^2+z_s^2}\right) \, .
\end{eqnarray}
where $\mathcal{K}$ is the complete elliptic integral of the first kind. This integral can be computed numerically.

\begin{figure}[t]
\centering
\includegraphics[width=0.5\textwidth]{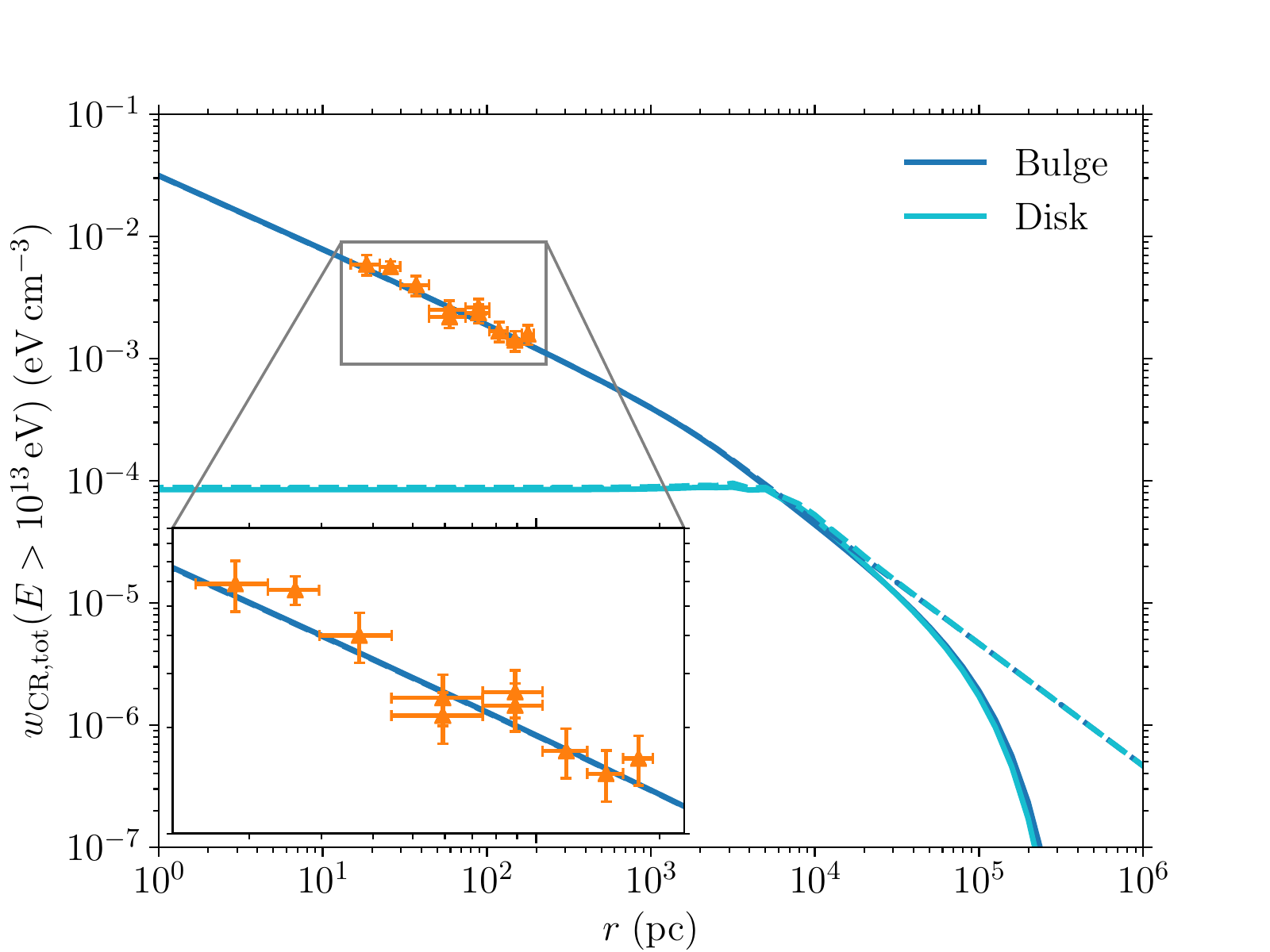}
\caption{ Total cosmic-ray density profiles for the bulge and disk populations of MSP, whether or not neglecting the error function component in Eq.~(\ref{eq:CRdens_HESS}) (respectively, dashed and solid), for $E > 10^{13}\,{\rm eV}$ and a continuous injection time $T \sim 3\,$Gyrs.
The injection parameters have been chosen to enable a comparison with the cosmic-ray densities derived in H.E.S.S. (orange points), where specific assumptions are made, see text. The vertical error bars correspond to $1\sigma$ confidence levels and the horizontal ones to the bin sizes.  A population in the disk alone fails to reproduce the observed CR distribution.}\label{Fig:CR_Dist}
\end{figure}

The total cosmic-ray densities for the two different populations are illustrated in Fig.~\ref{Fig:CR_Dist}. Note that in this section, we only aim at comparing the shape of the radial cosmic-ray density profile and not its normalization. As we will see in the next section, a more realistic cosmic-ray injection from MSP is needed to determine the pulsar population parameters required to reproduce the data. These parameters only impact the normalization of the profile and not its general shape. Moreover, the cosmic-ray density derived from the H.E.S.S. measurements \cite{HESS_GC16} displayed in Fig.~\ref{Fig:CR_Dist} are obtained under different assumptions than ours. The luminosity of several regions is associated with the cosmic-ray density, using in particular the mass estimate in each region is based on tracer molecules.

For distances $r < 200 \,{\rm pc}$, the disk component is characterised by a constant cosmic-ray density profile. More detail about the disk cosmic-ray density is given in Appendix~\ref{App:CR_dist_study}. Hence we can readily see that the disk population alone cannot be sufficient to reproduce the results obtained in Ref.~\cite{HESS_GC16}, and that a bulge component is needed. Interestingly, the spatial distribution of the bulge MSP population allows to reproduce the radial dependency of the CR densities derived in Ref.~\cite{HESS_GC16}.

\section{Diffuse gamma-ray emission}\label{sec:flux}
In order to compute the diffuse gamma-ray flux associated with the total cosmic-ray densities, we first need to give more details on the cosmic-ray injection from MSP. As stated before, energetic particles are continuously injected for a typical duration $t_{\rm sd} = 9Ic^3P^2 / 8\pi^2B^2R_\star^6$ --the so-called spin-down timescale \cite{Shapiro83}.  A {\it transient} flux of cosmic-rays can be naturally modeled assuming that the electromagnetic energy of the pulsar wind, stemming from the combination of the stellar rotation and dipole magnetic field, is dissipated at each instant into particles. Following the notations and assumptions used in Section~\ref{sec:mspulsars}, the flux can be characterized by a mono-energetic injection at each time $t$ with energy $E_{\rm CR}(t) = E_0 (1+t/t_{\rm sd})^{-1}$. This type of injection produces a hard injection spectrum in $E^{-1}$. However, this first injection can be reprocessed, for instance at a shock front, producing a power-law injection spectrum with a possibly softer index, if the acceleration process is stochastic for instance. This flux can then be modeled as a {\it uniform} power-law spectrum that lasts over $t_{\rm sd}$.

The total cosmic-ray density is calculated by accounting for the spatial distribution of MSP and diffusion of cosmic-rays (see Section~\ref{sec:diff}) but also for the variety of MSP in the population considered. We model the initial spin period distribution by a log-normal $F_P[P({\rm ms})] \propto P^{-1} \exp[(\log P-\mu)^2/2\sigma ^2]$  with $\mu = 1.5 $ and $\sigma = 0.58$  \citep{Lorimer15}, and the magnetic field distribution by a power-law $F_B(B) \propto B^{-1}$ for $B_{\rm min}<B<B_{\rm max}$ \citep{Story07}, where we set $B_{\rm min} = 10^8\,{\rm G}$ and $B_{\rm max} = 10^{11}\,{\rm G}$. Higher values of $B_{\rm max}$ also allow to reproduce the H.E.S.S. observations but require a larger number of pulsars. Lower values of $B_{\rm max}$ do not allow a fit to the H.E.S.S. data at the highest energies.

The predictions for the gamma-ray diffuse emission can be compared with the H.E.S.S. observations \cite{HESS_GC16}. The gamma-ray diffuse flux and integrated luminosity are measured in different regions close to the GC.
Considering that the gamma-ray emission is entirely produced by $pp$ interactions, we can calculate the gamma-ray spectrum ${\rm d}N_\gamma/{\rm d}\epsilon {\rm d}t$, where $\epsilon$ is the photon energy, from the differential cross section for the gamma-ray production ${\rm d}\sigma_{pp,\gamma}(\epsilon,E)/{\rm d}\epsilon$, the cosmic-ray density $w_{\rm CR}(E,r)$ and the mass of the target $M$ in the region of interest, centred at $r$ (where we consider that $w_{\rm CR}$ is constant):
\begin{eqnarray}\label{eq:ngamma}
\frac{{\rm d}N_\gamma}{{\rm d}\epsilon {\rm d}t} &=& \eta_{\rm N}\int {\rm d}E \,\frac{{\rm d}N_p}{{\rm d}E {\rm d}t} \frac{{\rm d}\sigma_{pp,\gamma}}{{\rm d}\epsilon}(\epsilon,E) \,, \nonumber \\
&=& \frac{\eta_{\rm N} Mc }{m_p} \int {\rm d}E \, w_{\rm CR}(E,r)\frac{{\rm d}\sigma_{pp,\gamma}}{{\rm d}\epsilon}(\epsilon,E) \,.
\end{eqnarray}
The factor $\eta_{\rm N}$ accounts for the presence of nuclei ($Z>1$) in interstellar matter and $m_p$ is the proton mass. The differential cross sections for the gamma-ray production are generated with the EPOS LHC model \cite{Werner06,Pierog:2013ria} and are illustrated in Appendix~\ref{App:dsigma}. We note that \cite{HESS_GC16} calculate gamma-ray luminosity as follows: $L_\gamma(\epsilon) \sim \eta_{\rm N} w_{\rm CR}(10 \epsilon) M / n_H \, m_p \, t_{pp \rightarrow \gamma}$, where $n_H$ is the hydrogen gas density and $t_{pp \rightarrow \gamma}$ is the proton energy-loss timescale related to gamma ray production. In our work, we integrate over the differential cross section for the gamma-ray production. Therefore, each proton energy is not related to a unique photon energy but to a distribution of photon energies, which is characterized by the differential cross section. For a monoenergetic proton injection at the energy $E$, the peak of the gamma-ray spectrum $\epsilon^2 {\rm d}N_\gamma/{\rm d}\epsilon$ is located around  $\epsilon \approx E /10$.

We can compare our predictions in the inner $50\,{\rm pc}$ region for the transient and uniform injection models, with H.E.S.S. measurements extracted from Ref.~\cite{HESS_GC16}, and Fermi-LAT data extracted from Ref.~\cite{Gaggero17}. The diffuse gamma-ray flux in this region is obtained from the gamma-ray luminosity: $\epsilon^2 \Phi_\gamma (\epsilon) = L_\gamma (\epsilon) / 4\pi D_{\rm GC}^2 \, \Delta \Omega $, where $D_{\rm GC} \sim 8 \times 10^3\,{\rm pc}$ is the distance from the galactic center and $\Delta \Omega \simeq \Delta\phi (\cos \theta_{\rm min}-\cos\theta_{\rm max})$ is the solid angle of the inner 50 pc region.

The uniform injection case, with a power-law injection, a constant maximum acceleration energy and luminosity over the pulsar spin-down time, is first treated in Section~\ref{Sec:cont}. The transient injection is then studied in more detail in Section~\ref{Sec:trans}. 
 
\subsection{Uniform power-law cosmic-ray injection}\label{Sec:cont}

The uniform injection of accelerated protons from each millisecond-pulsar $\dot{Q}_p(E)$ is modeled by a simple power-law ${\rm d}N/{\rm d}E \propto E^{-\beta}$ for $E_{p,\rm min}<E<E_{p,\rm max}$, with $\beta$ the injection spectral index. Each pulsar injects protons during the typical duration $T = t_{\rm sd}$. We obtain the following proton injection rate 
\begin{equation}
\dot{Q}_p(E) = \frac{\eta_p L_{\rm CR}(t_{\rm sd}) (2 - \beta)}{1-[E_{p, \rm min}/E_{p, \rm max}(t_{\rm sd}) ]^{2-\beta}} \left[\frac{E}{E_{p, \rm max}(t_{\rm sd}) }\right]^{2-\beta}  \, ,
\end{equation}
where $\eta_p$ is the baryon loading, $L_{\rm CR}(t_{\rm sd}) \sim 2.5 \times 10^{35}\,{\rm erg\,s}^{-1}\, \eta_{\rm acc} \, B_{9}^{2} \, R_{\star,6}^{6} \, P_{{\rm i},-3}^{-4}$ is the pulsar luminosity in cosmic rays at $t_{\rm sd} \sim 9.8\times 10^{7}\,{\rm yr}\,I_{45}B_{9}^{-2}R_{\star,6}^{-6}P_{{\rm i},-3}^2$ and $E_{p,\rm max}(t_{\rm sd}) \sim 2.3 \times 10^{14}\,{\rm eV} A\,\eta_{\rm acc}\,\kappa_4^{-1} B_{9} R_{\star,6}^3 P_{{\rm i},-3}^{-2}$ is the maximum energy of accelerated protons at $t_{\rm sd}$. In this case study, the cosmic-ray luminosity and maximum energy do not vary, and we choose the fiducial minimum injection energy $E_{p,\rm min} = 10^{10}\,{\rm eV} $. If we neglect the impact of spallation on the diffusion, and assume a constant pulsar birth rate $\tau_{\rm birth}$ during the time $t$, the cosmic-ray density for one source is given by Eq.~(\ref{eq:CRdens_HESS}). For each sub-class of pulsars with fixed $P$ and $B$, the cosmic-ray density is weighted by $t_{\rm sd}/t$, ensuring a uniform emission during the time $t$. This formalism is valid as long as the average timescale between two millisecond pulsar births is shorter than the spin-down timescale $1/\tau_{\rm birth}\ll t_{\rm sd}$, which is usually the case, as $t_{\rm sd}\gtrsim 10^7\,$yrs for MSP, and the typical birth rate is $\tau_{\rm birth} \gtrsim 1/345000 \,{\rm yr}^{-1}$ \cite{Ferrario07,Story07,Lorimer2008}. We calculate the cosmic-ray density integrated over the spin and magnetic field distributions, illustrated in Appendix.~\ref{App:CR_dens_BP}. Without the additional factor $t_{\rm sd}/t$, the cosmic-ray density spectrum would be well described by a power-law of index $\beta + \delta$, but in our case the situation is more complex. The cosmic-ray density as a function of distance is still well described by a power-law $\propto r^{-1}$, as it is not influenced by the integration over the distributions.

\begin{figure}[t]
\centering
\includegraphics[width=0.49\textwidth]{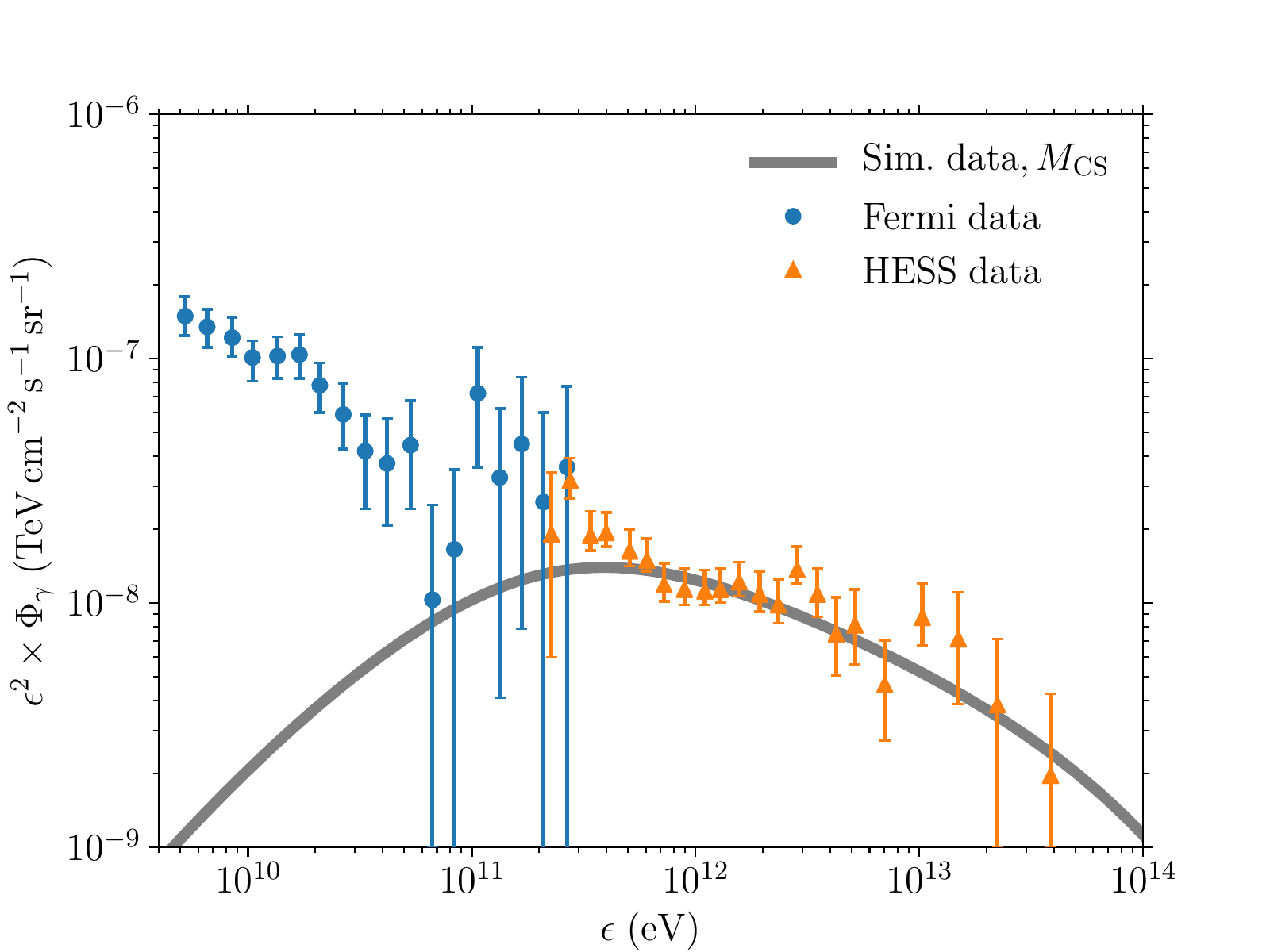}
\includegraphics[width=0.49\textwidth]{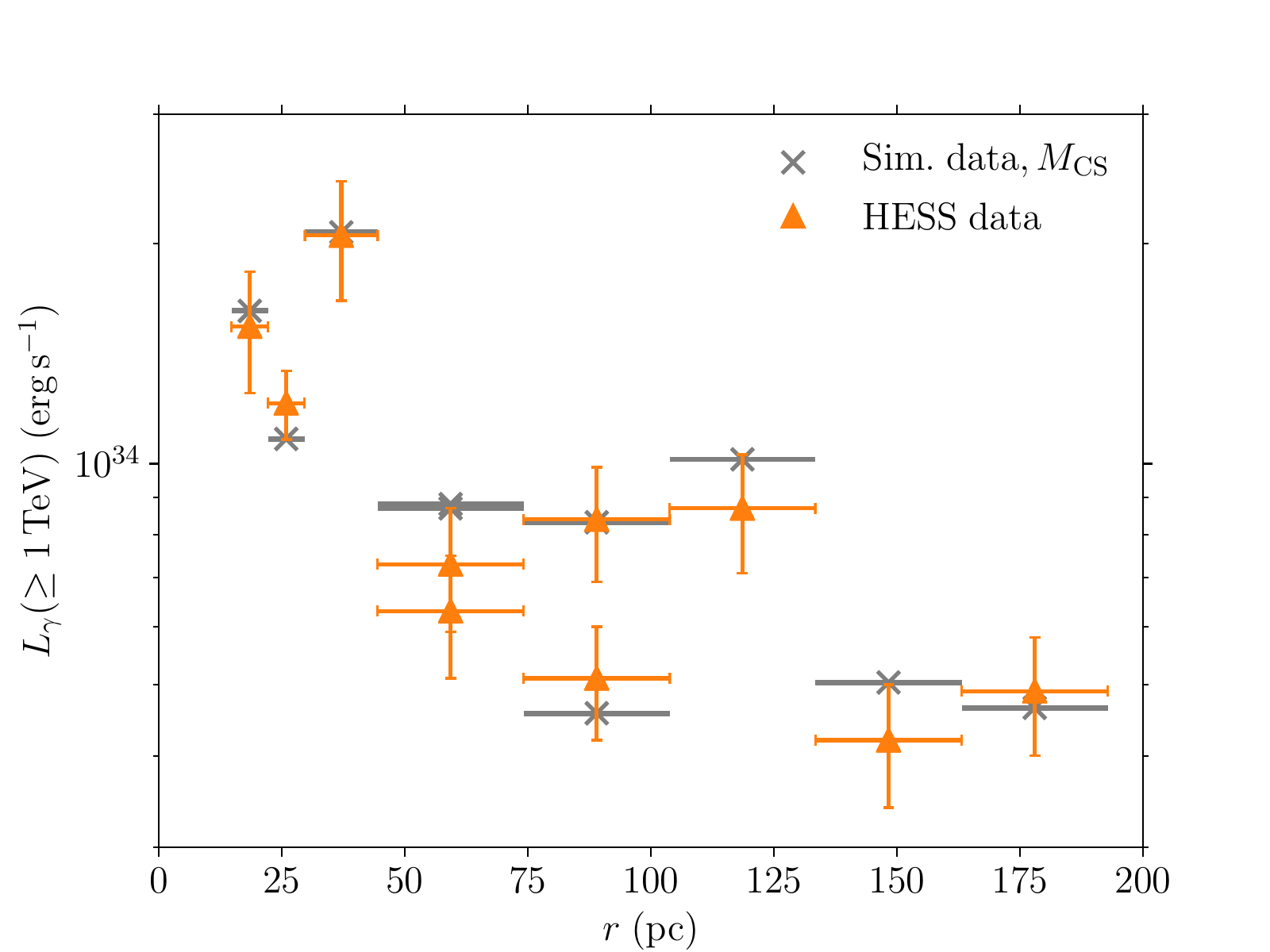}
\caption{Diffuse gamma-ray spectrum in the central region ({\it left}), as predicted by our uniform power-law injection model for $\beta = 1.1$, $E_{p,\rm min} = 10^{10}\,{\rm eV} $, $\eta_{\rm acc} \sim 0.03$ and $\eta_p N_{\rm b} \simeq 2 \times 10^6$ (grey thick line) and measured by Fermi (blue dots) and H.E.S.S. (orange triangles). Gamma-ray luminosity as a function of the distance to the galactic center ({\it right}), from our model with the same parameters (grey crosses) and measured by H.E.S.S. (orange triangles). The horizontal bars show the bin size and the vertical ones, the 1$\sigma$ confidence level of the H.E.S.S. data.}\label{Fig:Gamma_CR_cont}
\end{figure}

The total cosmic-ray injection is obtained after the integration over the spatial distribution of the bulge MSP population, which gives the dominant contribution for $N_{\rm d} \lesssim 10 N_{\rm b}$ in the region of interest $r \lesssim 200\,{\rm pc}$. As $w_{\rm CR}(E,r,t) \propto r^{-1}$ for $r \lesssim 200\,{\rm pc}$, we can use the analytical expressions derived in Section~\ref{sec:diff}, especially the equation~\ref{Eq:wCR_int_dist}. Finally, we calculate the diffuse gamma-ray spectrum and luminosity,  illustrated on Figure~\ref{Fig:Gamma_CR_cont} for  $\beta = 1.1$, $E_{p,\rm min} = 10^{10}\,{\rm eV} $, $\eta_{\rm acc} \sim 0.03$ and $\eta_p N_{\rm b} \sim 10^6$. We recall that $\eta_{\rm acc}$ is obtained for a pair multiplicity $\kappa=10^3$. In general the H.E.S.S. data can be reproduced for $\eta_{\rm acc}\,(m_p/2m_e\kappa)/(1+m_p/2m_e\kappa) \sim 10^{-2}$. We also note that $N_{\rm b}$ corresponds to the total number of MSP in the bulge population. A better quantity to compare with other MSP pulsation studies would be the number of pulsars with cosmic-ray luminosities lying in a given range. The population analysis in the literature consider frequently luminosities $>L_{\gamma,{\rm min}}\sim10^{33}\,{\rm erg\,s}^{-1}$ for gamma rays produced through leptonic processes. Using the distributions calculated in Appendix~\ref{App:CR_dens_BP}, we can estimate that a sub-population of MSP with $L_{\rm CR}(t_{\rm sd})>10^{33}\,{\rm erg\,s}^{-1}$ represents $\sim 10\%$ of the total MSP population, for $\eta_{\rm acc} = 0.03$ and $\kappa = 10^3$. The corresponding number of MSP in this sub-population is then $N_{\rm b}({L_{\rm CR}(t_{\rm sd})>10^{33}\,{\rm erg\,s^{-1}}}) \sim  10^5$. Note that a higher lower bound for $L_{\rm CR}(t_{\rm sd})$ would lead to a smaller fraction of the total population. Our lower bound $L_{\rm CR}(t_{\rm sd}) = 10^{33}\,{\rm erg\,s}^{-1}$ is conservative in this sense.

This model allows us  to fit  the data in the H.E.S.S. energy range. However, the fact that for each pulsar, the cosmic-ray maximum energy and luminosity do not vary, make this model quite unrealistic. Moreover, the value of the parameter $E_{p,\rm min} = 10^{10}\,{\rm eV} $ is quite arbitrary, but determines the energy range covered by the modeled diffuse gamma-ray spectrum.

We note that hard injection spectra $\beta\sim 1-2$ are needed to fit the H.E.S.S. data. Such spectra can be achieved in pulsars, for example via reconnection processes in the striped wind, as shown by hybrid and particle-in-cell simulations (e.g., \cite{Bennett95,Dieckmann09,Spitkovsky08,Sironi11}). More simply, the unipolar induction toy-model in the transient monoenergetic injection scenario described in the next section produces naturally hard injection spectra with $\beta=1$ \cite{Shapiro83}, without involving additional parameters as the minimum injection energy.

\subsection{Transient monoenergetic cosmic-ray injection}\label{Sec:trans}

From Eq.~(\ref{eq:LCR}), the transient flux of cosmic rays injected into the wind is given by
\begin{equation}
\frac{{\rm d}^2N}{{\rm d}E{\rm d}t}(E,t) = \frac{9}{4}\frac{c^2 I}{ZeBR_\star^3} \, E^{-1}(t+t_{\rm sd})^{-1} \, ,
\end{equation}
with a mono-energetic injection at each time $t$ at $E_{\rm CR}(t) = E_0 (1+t/t_{\rm sd})^{-1}$, where $t_{\rm sd} \sim 3.1\times 10^{15}\,{\rm s}\,I_{45}B_{9}^{-2}R_{\star,6}^{-6}P_{{\rm i},-3}^2$. Following the approach of \cite{Blasi12a}, we calculate the CR density at position $\vec{r}$, energy $E$ and time $t$, for a transient CR injection from a single millisecond-pulsar located at $\vec{r}_s$ and starting at $t_s$
\begin{eqnarray}
w_{\rm CR}(E,\vec{r},t) &=& \int_{t^\star=0}^{t-t_s} {\rm d}t^\star \, E^2 \frac{{\rm d}^2N}{{\rm d}E{\rm d}t}(E,t^\star)\,{\cal G}(\vec r, t; \vec r_{s},t^\star) \, ,\\
&=& \int_{t^\star=0}^{t-t_s} {\rm d}t^\star \frac{9}{4}\frac{c^2 I}{ZeBR_\star^3}\,E (t^\star+t_{\rm sd})^{-1} \, \delta \left(\frac{E(1+t^\star/t_{\rm sd})}{E_0}-1 \right) \,{\cal G}(\vec r, t; \vec r_{s},t^\star) \, , \nonumber \\
&=& \frac{9}{4}\frac{ c^2 I E}{ZeBR_\star^3}\,{\cal G}(\vec r, t; \vec r_{s},t_s+t_{\rm sd}(E_0/E-1))  \, , \nonumber
\end{eqnarray}
which is non zero only for $E_0 [1+(t-t_s)/t_{\rm sd}]^{-1} \leq E \leq E_0$. For a given energy $E$, this solution is only valid after the injection of cosmic-rays, for $t > t_s + t_{\rm sd}(E_0/E-1)$.

We assume a uniform distribution for the birth time of the pulsars $t_s$ between $t_s=0$ and $T_{s,{\rm max}} = t$. We integrate the one-source cosmic-ray density over the birth time, spin and magnetic field distributions
\begin{eqnarray}
w_{\rm CR}(E,\vec{r},t) &=&   \int_{B = 0}^\infty {\rm d} B \int_{P = 0}^\infty {\rm d}P \int_{t_s=0}^{t_{s,\rm max}} {\rm d}t_s \, \frac{9}{4}\frac{ c^2 I E F_{B}(B) F_{P}(P)}{eBR_\star^3 \, T_{s,{\rm max}}}  {\cal G}[\vec r, t; \vec r_{s},t_s+t_{\rm sd}(E_0/E-1)] \nonumber \\
&=&   \int_{B = 0}^\infty {\rm d} B \int_{P = 0}^\infty {\rm d}P \int_{t_s'=t'_{s,{\rm min}}}^{t'_{s,{\rm max}}} {\rm d}t'_s \, \frac{9}{4}\frac{ c^2 I E F_{B}(B) F_{P}(P)}{eBR_\star^3 \, T_{s,{\rm max}}} \, {\cal G}(\vec r, t'_s; \vec r_{s},0) \, ,
\end{eqnarray}
with $t_{s, \rm max}=\min\{\max[t- t_{\rm sd}(E_0/E-1),0],T_{s,{\rm max}}\}$, $t'_{s,{\rm min}}= t- t_{\rm sd}(E_0/E-1)-t_{s, \rm max}$ and $t'_{s,{\rm max}} = t- t_{\rm sd}(E_0/E-1)$. Note that $t_{\rm sd}$ and $E_0$ depend on $B$ and $P_{\rm i}$. Different cosmic-ray densities accounting for the $P$, $B$ and $t_s$ distributions are illustrated in Appendiz~\ref{App:CR_dens_BP}.

For a given set of MSP parameters $(P_{\rm i},B,R_\star,t_s)$, the maximum and minimum energies of cosmic-rays can be very close for a small observation time $t>t_s$, as $E_{\rm max}=E_0$ and $E_{\rm min}=E_0 \{1+[\min(t,T_{s,{\rm max}})-t_s]/t_{\rm sd}\}^{-1}$. For $t_{\rm obs}>T_{s,{\rm max}}$, the cosmic-ray density is attenuated very rapidly. In this study, we focus on the case $t_{\rm obs} \leq T_{s,{\rm max}}$, considering that the birth of MSP in the Galactic center still arise today. The value of $T_{s,{\rm max}}$ influences the normalization of the birth time distribution and should therefore be chosen carefully; from the typical age of our galaxy, we set $T_{s,{\rm max}} = 10^{17}\,{\rm s}$.

As shown in Section~\ref{sec:CR_dens}, for $r\lesssim100\,{\rm pc}$, the cosmic-ray density as a function of distance $r$ is well described by a power-law $\propto r^{-1}$. Therefore, the total cosmic-ray density can be obtained by using the results of Section~\ref{Sec:Total_CR_dens}, which accounts for the integration of the above density over the spatial distribution of MSP in the bulge. As stated before, the contribution of the bulge population is dominant for $N_{\rm d} \lesssim 10 N_{\rm b}$ at $r \lesssim 200\,{\rm pc}$. Therefore, depending on the relative number of MSP in both populations, the disk population could contribute to the diffuse flux: for $N_{\rm d}/N_{\rm b} \sim 1$, it would give a significant contribution above $r\sim 5 \times 10^3\,{\rm pc}$ (see Fig.~\ref{Fig:CR_Dist}).

\begin{figure}[t]
\centering
\includegraphics[width=0.49\textwidth]{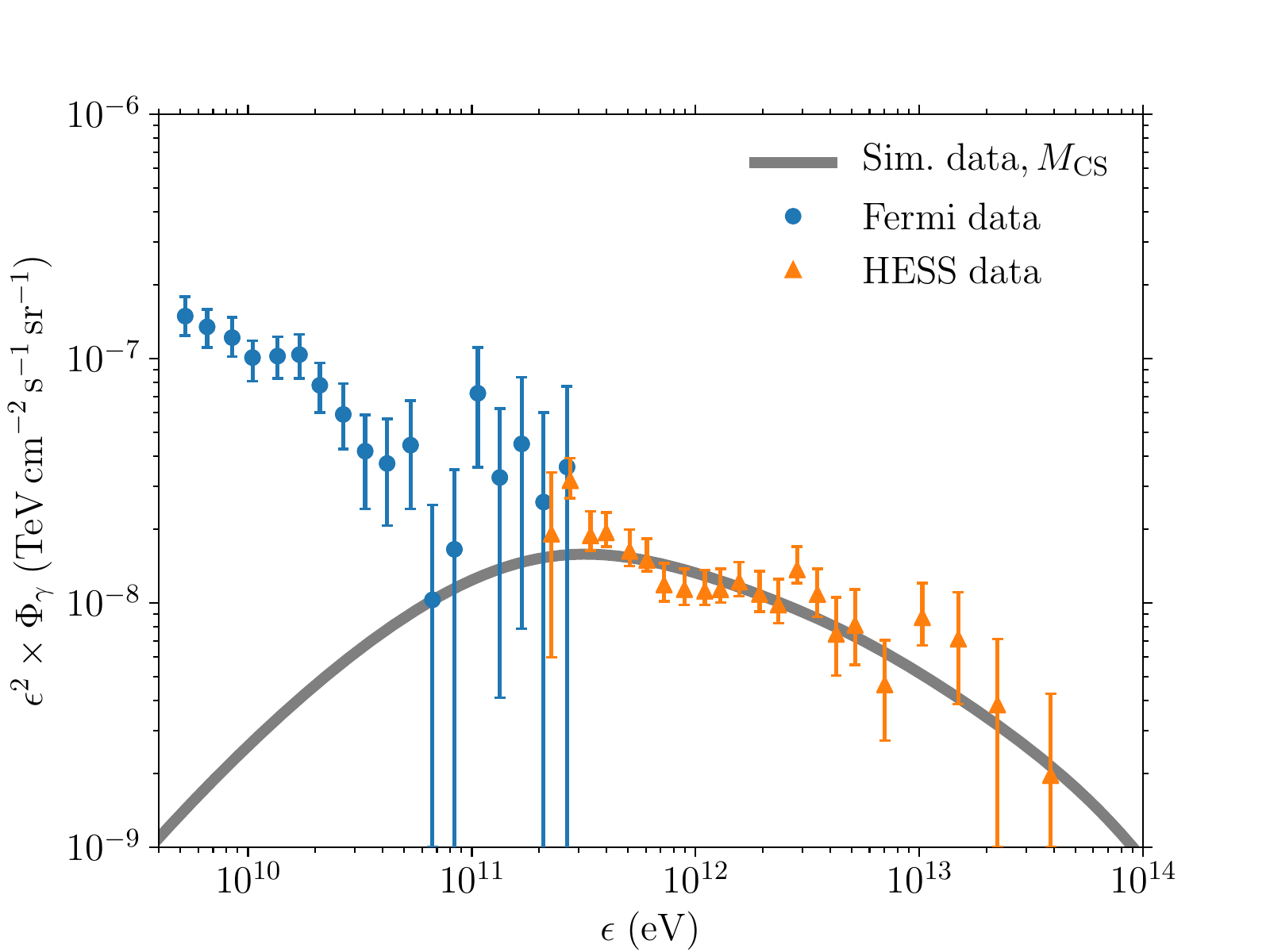}
\includegraphics[width=0.49\textwidth]{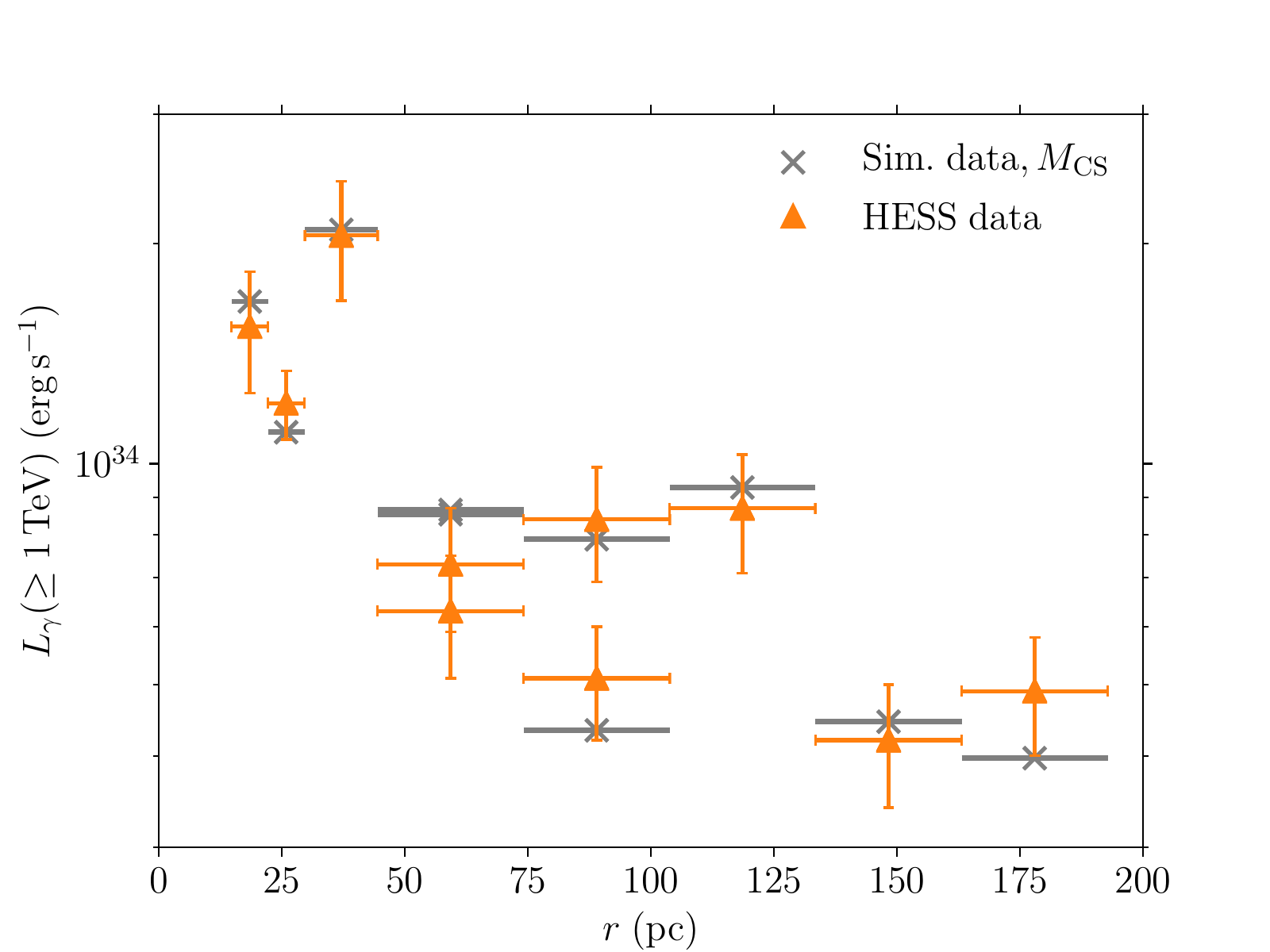}
\caption{Diffuse gamma-ray spectrum in the central region ({\it left}), as predicted by our transient monoenergetic injection model for a mass estimate based on CS tracers, $\eta = 0.03$,  $\eta_p N_{\rm b} \simeq 10^6$ (grey thick line) and measured by Fermi (blue dots) and H.E.S.S. (orange triangles). Gamma-ray luminosity as a function of the distance to the galactic center ({\it right}), from our model with the same parameters (grey crosses) and measured by H.E.S.S.  (orange triangles).}\label{Fig:Gamma_CR}
\end{figure}

Considering the bulge contribution only, for a mass estimate based on CS tracers, a moderate acceleration efficiency $\eta_{\rm acc} \sim 0.03$ for $\kappa = 10^3$, a total number of pulsars $\eta_p N_{\rm b} \sim 10^6$, $\eta_{\rm N} = 1.5$ and a power-law distribution of the magnetic field of index $-1$ between $B_{\rm min} = 10^8\,{\rm G}$ and $B_{\rm max} = 10^{11}\,{\rm G}$, we obtain a gamma-ray spectrum and a luminosity profile that are compatible with the H.E.S.S. measurements. 
Our results are shown in Figure~\ref{Fig:Gamma_CR}. The $1/B$ dependence of the magnetic field distribution is required to obtain the correct power-law shape of the cosmic-ray density and diffuse gamma-ray spectra, and therefore a good match to the H.E.S.S. measurements. As explained in Section~\ref{Sec:cont}, considering the number of MSP in a sub-population with cosmic-ray luminosities $L_{\rm CR}(t_{\rm sd}) > 10^{33}\,{\rm erg\,s}^{-1}$ can be of interest for the comparison with other MSP population studies. In the case of a transient cosmic-ray injection, we obtain $N_{\rm b}({L_{\rm CR}(t_{\rm sd})>10^{33}\,{\rm erg\,s^{-1}}}) \sim 7 \times 10^4$. Interestingly, the MSP number with $L_{\rm CR}(t_{\rm sd}) \ge 10^{34}\,{\rm erg\,s}^{-1}$ is found to be of the same order that the one derived in Ref.~\citep{Ploeg17}.

\section{Discussion and conclusions}\label{sec:discussion}

A total population of $\eta_p N_{\rm b} \sim 10^6$ millisecond pulsars (MSP), accelerating protons up to very high energies with baryon loading $\eta_p$, appears as an acceptable candidate to explain the diffuse gamma-ray excess observed by H.E.S.S. in the Galactic center region. Regarding the properties of these pulsars, moderate acceleration efficiencies $\eta_{\rm acc} \sim 0.03$ with pair multiplicities $\kappa = 10^3$, specific initial spin and dipole magnetic field distributions are required to reproduce the spectral and spatial characteristics of the data. We note that $\eta_{\rm acc}$ and $\kappa$ are interlinked/correlated parameters. The pulsar population considered is located in a $10^3\,{\rm pc}$ bulge around the Galactic center. The ratio between the number of pulsars in the disk and the bulge components $N_{\rm d}/N_{\rm b}$ should be smaller than $\sim 10$, so that the bulge component remains predominant.

The contribution of heavier nuclei appears only as a pre-factor $\eta_N$ in the gamma-ray luminosity calculation. This value is commonly chosen to be $\eta_N=1.5$, e.g., \cite{HESS_GC16}. However, a more refined treatment would be required to account for the various and more complex effects appearing if pulsars accelerate protons as well as heavier nuclei. For instance the accelerated nuclei would reach energies higher than protons, as typically $E_{N,{\rm max}} \sim Z E_{p,{\rm max}}$. The spallation of nuclei would also create secondary nuclei during the diffusion process. Such effects are left for future studies.

We have modeled the diffusion process of cosmic rays using a standard diffusion coefficient \cite{GALPROP}. Recent detections of extended TeV emissions around young pulsars with HAWC has led to a measurement of the diffusion coefficient, that the collaboration claims as a general value for the interstellar medium \cite{Abeysekara17}. Reference~\cite{Hooper17} however argues that this measurement should be only valid locally, around the Geminga and Monogem pulsars. Note also that the statistical significance of these measurements is still low and to be confirmed. We tested the influence of the diffusion coefficient on our results: for a $100$-times lower diffusion coefficient, the radial extension of the gamma-ray excess is reduced, and therefore the gamma-ray luminosity as a function of the distance to the galactic center cannot match with the H.E.S.S. observations. This result is illustrated in Appendix~\ref{App:D_impact}.

The modelling of our population of MSP is subject to uncertainties. In particular, the dipole magnetic field distribution of such objects is still not well constrained by the observations. However, we noted in this work that this distribution has a strong impact on the predictions, especially on the shape of the gamma-ray spectrum. Whereas $B_{\rm max}$ has a minor impact as long as $B_{\rm max} \geq 10^{11}\,{\rm G}$, the index $-1$ of this power-law distribution is decisive in order to match the H.E.S.S. measurements.

We focussed in this work on the gamma-ray spectrum at the highest energies, {\it i.e.} $\sim 1\,{\rm TeV}$, and obtained a reasonable match to the diffuse emission measured by H.E.S.S. with our two cosmic-ray injection models, for a bulge population of MSP. However, these hadronic models do not account for the Fermi gamma-ray observations at lower energies. As mentioned in Section~\ref{section:GeVobs}, standard leptonic scenarios involving populations of pulsars can explain the flux observed by Fermi up to $\sim 10 \,{\rm GeV}$ energies. This would imply that the hadronic and leptonic emissions from MSP would fail to explain the gamma-ray flux observed by Fermi between $10 \,{\rm GeV}$ and $100 \,{\rm GeV}$, shown in Fig.~\ref{Fig:Gamma_CR_cont} and \ref{Fig:Gamma_CR}. On the other hand, several theoretical models have been discussed in the literature that enable the acceleration of electron and positron pairs in pulsars up to $\sim 10\,$TeV energies \cite{Kisaka12,Bednarek13,Venter15}. Observationally, two young pulsars (Geminga, Monogem) have been identified by HAWC as emitters of gamma rays up to 100\,GeV via leptonic components \cite{Abeysekara17}. Hints of TeV halos around MSP have been reported to be found in the HAWC data \cite{Hooper18}. Moreover, \cite{Petrovic15,Yuan15} suggest that the up-scattering of low-energy photons by electron-positron pairs emitted by the MSPs could contribute to the gamma-ray diffuse emission above a few GeV. These arguments can be invoked to suggest that the gamma-ray emission observed by Fermi could be explained up to $\sim 100\,$GeV by leptonic emission from MSP, and that the hadronic component would then take over. MSP would appear as the dominant sources of the gamma-ray diffuse emission at very high energies in the Galactic Center.

Nearby pulsars such Geminga and Monogem may contribute to the CR density at Earth. In particular, assuming a baryon loading  of 0.05 in the Geminga pulsar and the diffusion coefficient used in this study away from the inner tens of pc around Geminga, a qualitative estimate of the Geminga CR contribution at Earth of about 10$^{-9}$ eVcm$^{-3}$ from the electron luminosity derived in Ref.~\cite{Abeysekara17}. Assuming the value of the diffusion coefficient inferred in the inner tens of pc around Geminga to be valid on the Geminga-to-Earth spatial scale would drastically increase the CR contribution. However, given the electron energy losses, such a low diffusion coefficient would not enable one to measure electrons up to 20 TeV, for which local CR electron sources such as Geminga are natural
sources

In order to reproduce the H.E.S.S. measurements, a total number of pulsars in the bulge $\eta_pN_{\rm b} \sim 10^6$ is required. This number is subject to large uncertainties, as it depends on the baryon loading $\eta_p$ -- a poorly constrained quantity, on the acceleration efficiency and on the various distributions characterizing our pulsar population. Better observational constraints would be required to obtain a more accurate estimate of this quantity. Moreover, $N_{\rm b}$ should not be compared directly with the number of MSP derived in other MSP population studies, such as \cite{FermiGC17,Ploeg17}. Their number of MSP are frequently given for gamma-ray luminosities in a given range, with gamma rays produced through leptonic processes. From our cosmic-ray luminosity distribution, about $10\%$ of the total MSP population is characterized by $L_{\rm CR}(t_{\rm sd}) > 10^{33}\,{\rm erg\,s}^{-1}$. A higher value of the cosmic-ray luminosity lower bound $L_{\rm CR}(t_{\rm sd}) > 10^{34}\,{\rm erg\,s}^{-1}$ lead to a even lower fraction $\sim 3\%$ of the total MSP population, which gives a number of pulsars in this luminosity interval $N_{\rm b}(L_{\rm CR}(t_{\rm sd}) > 10^{34}\,{\rm erg\,s}^{-1}) \sim 10^4-10^5$, more compatible with the values obtained in \cite{FermiGC17,Ploeg17}. A more detailed treatment would require a comparison between our hadronic model and leptonic scenarii.

More precise measurements above $50\,{\rm TeV}$ using deeper observations of the GC region with H.E.S.S.  and future high-sensitivity instruments such as CTA, whether or not indicating the presence of a high energy cut-off in the VHE diffuse emission spectrum, would put strong constraints on several parameters of our model, as the acceleration efficiency $\eta_{\rm acc}$ or the magnetic field distribution --especially on its upper bound $B_{\rm max}$, the value of $B_{\rm min}$ being already better constrained by observations. A high energy cut-off would be associated with a low $\eta_{\rm acc}$ or a low $B_{\rm max}$.

\section*{Acknowledgements}
CG is supported by a fellowship from the CFM Foundation for Research and by the Labex ILP (reference ANR-10-LABX-63, ANR-11-IDEX-0004-02). KK, CG and JS are supported by the APACHE grant (ANR-16-CE31-0001) of the French Agence Nationale de la Recherche. 

\appendix

\section{Cosmic-ray densities for pulsar populations}\label{App:CR_dist_study}

\begin{figure}[t]
\centering
\includegraphics[width=0.5\textwidth]{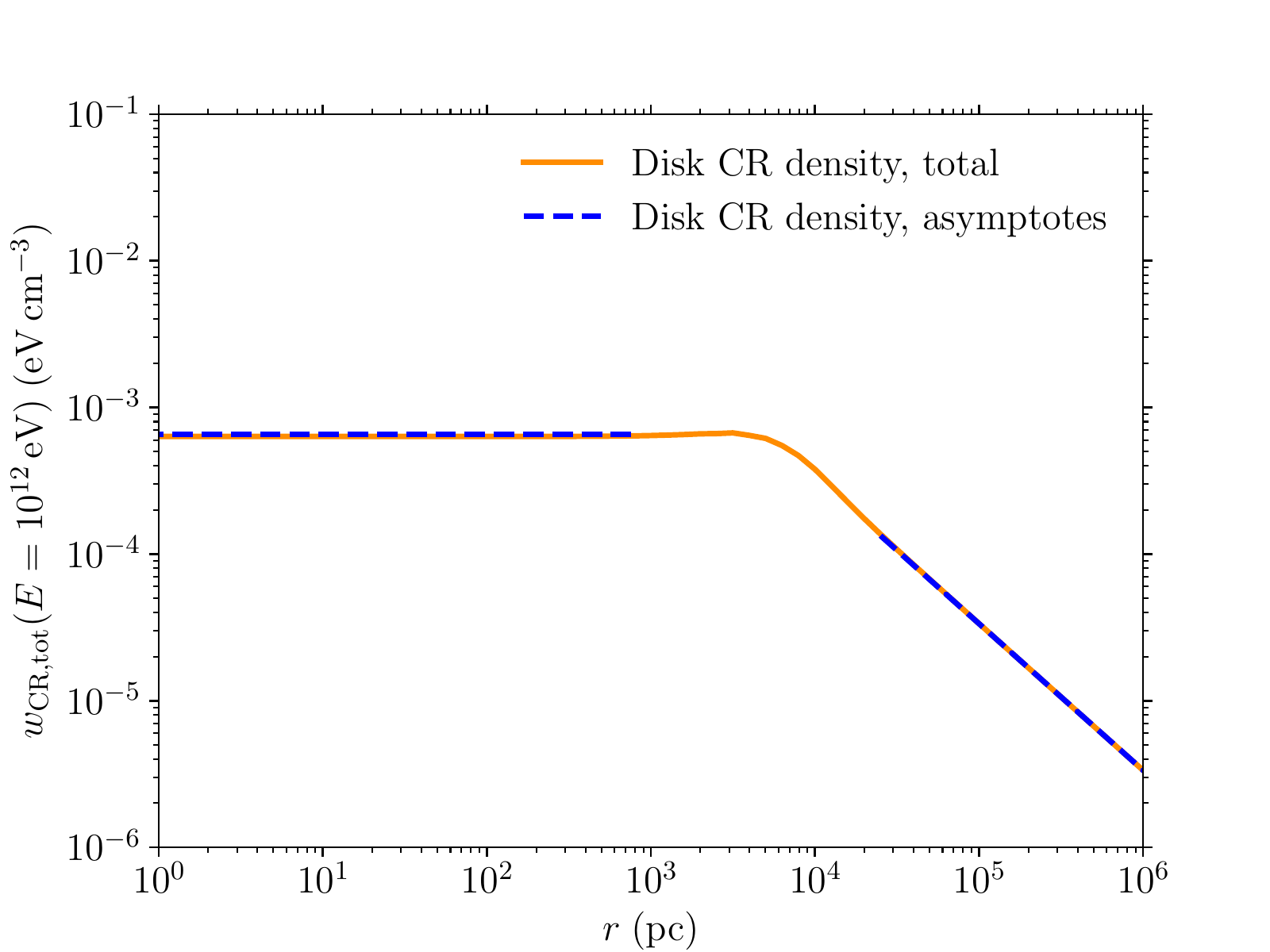}
\caption{Cosmic-ray density for the millisecond-pulsar disk population, with $\eta_{\rm acc} = 0.3$, $\beta = 2.2$, $t=10^{25}\,{\rm s}$, $N_{\rm b} = 1300$, $N_{\rm d} = 3500$ and $E = 10^{12}\,{\rm eV}$ (solid, blue), with asymptotes for low and high distances $r$ (dashed, blue).}\label{Fig:CR_Dens_Tot_lims}
\end{figure}

We study the integration over the pulsar populations, to obtain the total cosmic-ray density. The disk distribution peaks around $r_{\rm peak} \sim 10^4\,{\rm pc}$. We consider times such as ${\rm erfc}(|\vec{r}-\vec{r}_s|/r_{\rm diff}) \sim 1$: for $t=10^{20}\,{\rm s}$ and $E=10^{12}\,{\rm eV}$, $r_{\rm diff} \sim 10^6 \,{\rm pc}$. For $r \ll r_{\rm peak}$ (and $|\vec{r}-\vec{r}| \ll r_{\rm diff}$), at order zero
\begin{eqnarray}
w_{\rm CR,tot}(E,r,t) &\approx& \frac{\dot{Q}_p(E)\,N_{\rm disk}}{4\pi D(E) \sigma^{n+2} \Gamma(n+2)} \int_{r_s = 0}^\infty  {\rm d}r_s r_s^{n} \exp(-r_s/\sigma) \, ,  \nonumber \\
&\approx& \frac{\dot{Q}_p(E)\,N_{\rm disk}}{4\pi D(E) \sigma (n+1)}  \, .
\end{eqnarray}
For $r \gg r_{\rm peak}$ (and $|\vec{r}-\vec{r}| \ll r_{\rm diff}$), at order zero
\begin{eqnarray}
w_{\rm CR,tot}(E,r,t) &\approx& \frac{\dot{Q}_p(E)\,N_{\rm disk}}{4\pi D(E) \sigma^{n+2} \Gamma(n+2) r} \int_{r_s = 0}^\infty  {\rm d}r_s r_s^{n+1} \exp(-r_s/\sigma) \, ,  \nonumber \\
&\approx& \frac{\dot{Q}_p(E)\,N_{\rm disk}}{4\pi D(E) r}  \, .
\end{eqnarray}
These two limits are illustrated in Fig.~\ref{Fig:CR_Dens_Tot_lims}. Note that the limit obtained at $r \gg r_{\rm peak}$ is similar to the cosmic-ray density of the bulge population at $r \geq r_{\rm max}$.

\section{Cosmic-ray densities for initial spin and dipole magnetic field distributions}\label{App:CR_dens_BP}

\begin{figure}[ht]
\centering
\includegraphics[width=0.9\textwidth]{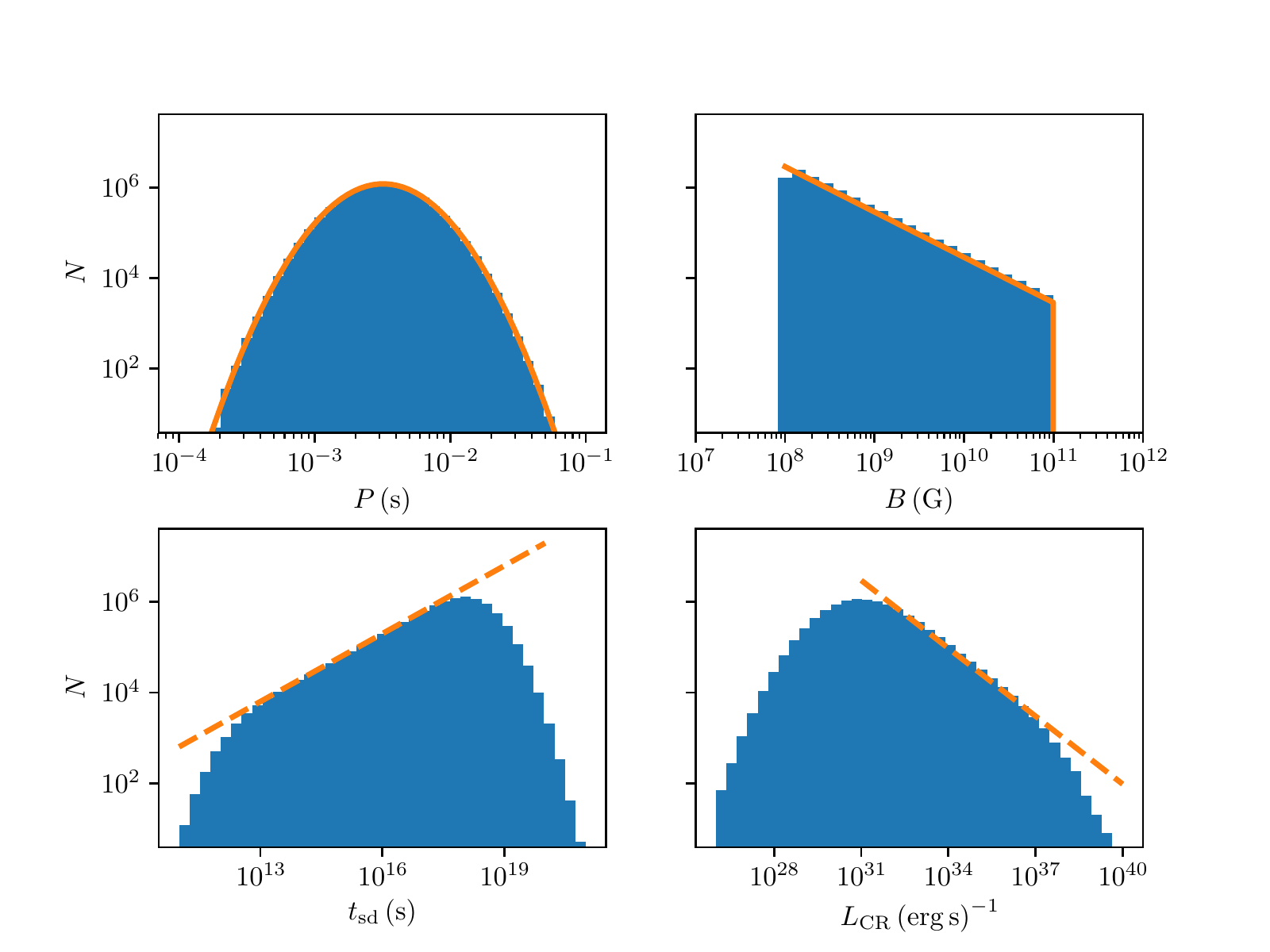}
\caption{From left to right and up to down, initial spin, dipole magnetic field, spin-down timescale and cosmic-ray luminosity distribution histograms, for a random draw of $10^7$ initial spin and dipole magnetic field values. The adjusted analytic distributions are also shown for initial spin and dipole magnetic field distributions (orange). }\label{Fig:Distributions}
\end{figure}

The initial spin and dipole magnetic field distributions are key ingredients allowing a realistic description of the millisecond pulsar population. These distributions, as well as the resulting spin-down timescale and cosmic-ray luminosity distributions, are illustrated in figure~\ref{Fig:Distributions} for $\eta_{\rm acc}=0.03$ and $\kappa = 10^3$. We see that the spin-down timescale distribution can be approximated by a power-law between  $\sim 10^{13}-10^{17}\,{\rm s}$, with an index $\sim 0.5$. The cosmic-ray luminosity distribution can also be approximated by a power-law between $\sim 10^{32}-10^{37}\,{\rm erg\,s}^{-1}$, with a soft profile $\propto L_{\rm CR}^{-0.5}$. This result is qualitatively compatible with the typical luminosity distribution of pulsars in gamma-ray, as the latter is well described by a harder profile.

\begin{figure}[ht]
\centering
\includegraphics[width=0.49\textwidth]{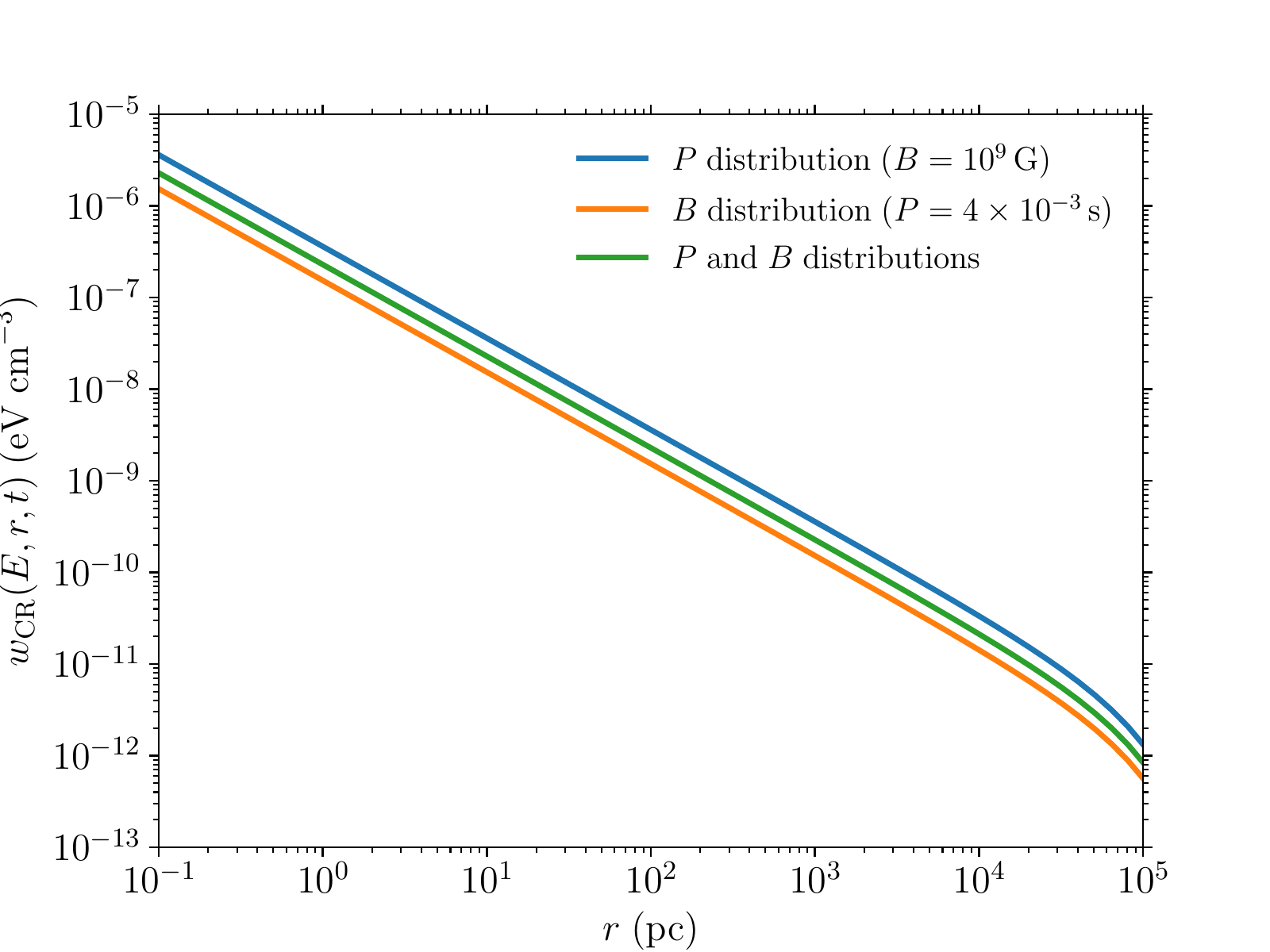}
\includegraphics[width=0.49\textwidth]{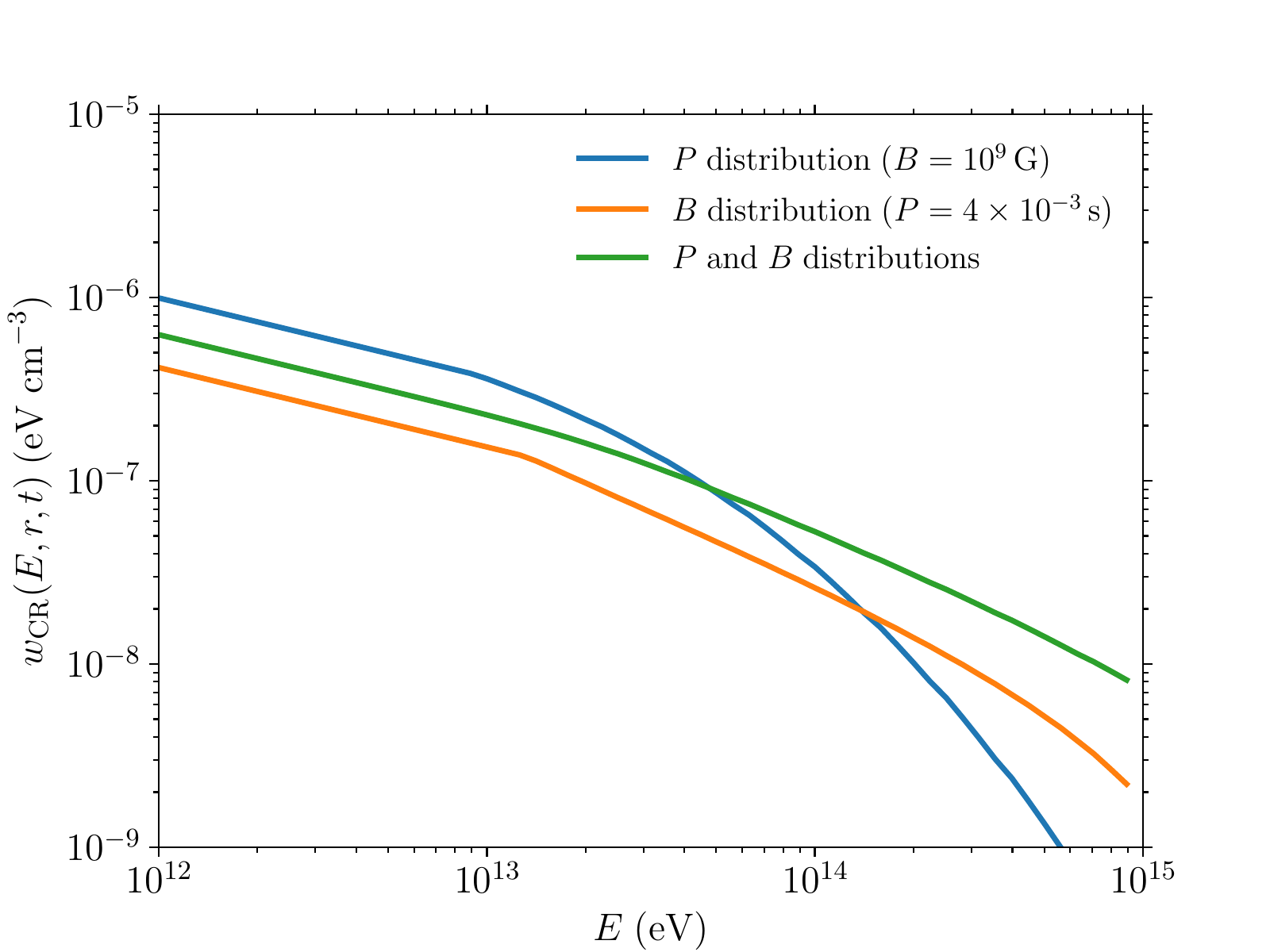}
\caption{Cosmic-ray density for $t\sim 3\,$Gyrs, $\beta = 2.1$, $\eta_{\rm acc} = 1$ and $E_{p,{\rm min}} = 10^{12}\,{\rm eV}$, as a function of distance for $E=10^{13}\,{\rm eV}$ ({\it left}) and as a function of energy for $r=1\,{\rm pc}$ ({\it right}). We show the cosmic-ray densities integrated on: $P$ distribution, for $B= 10^{9}\,{\rm G}$ (blue);  $B$ distribution, for $P=4\times10^{-3}\,{\rm s}$ (orange); $P$ and $B$ distributions (green).}\label{Fig:Int_PBt_dist_const}
\end{figure}

\begin{figure}[ht]
\centering
\includegraphics[width=0.49\textwidth]{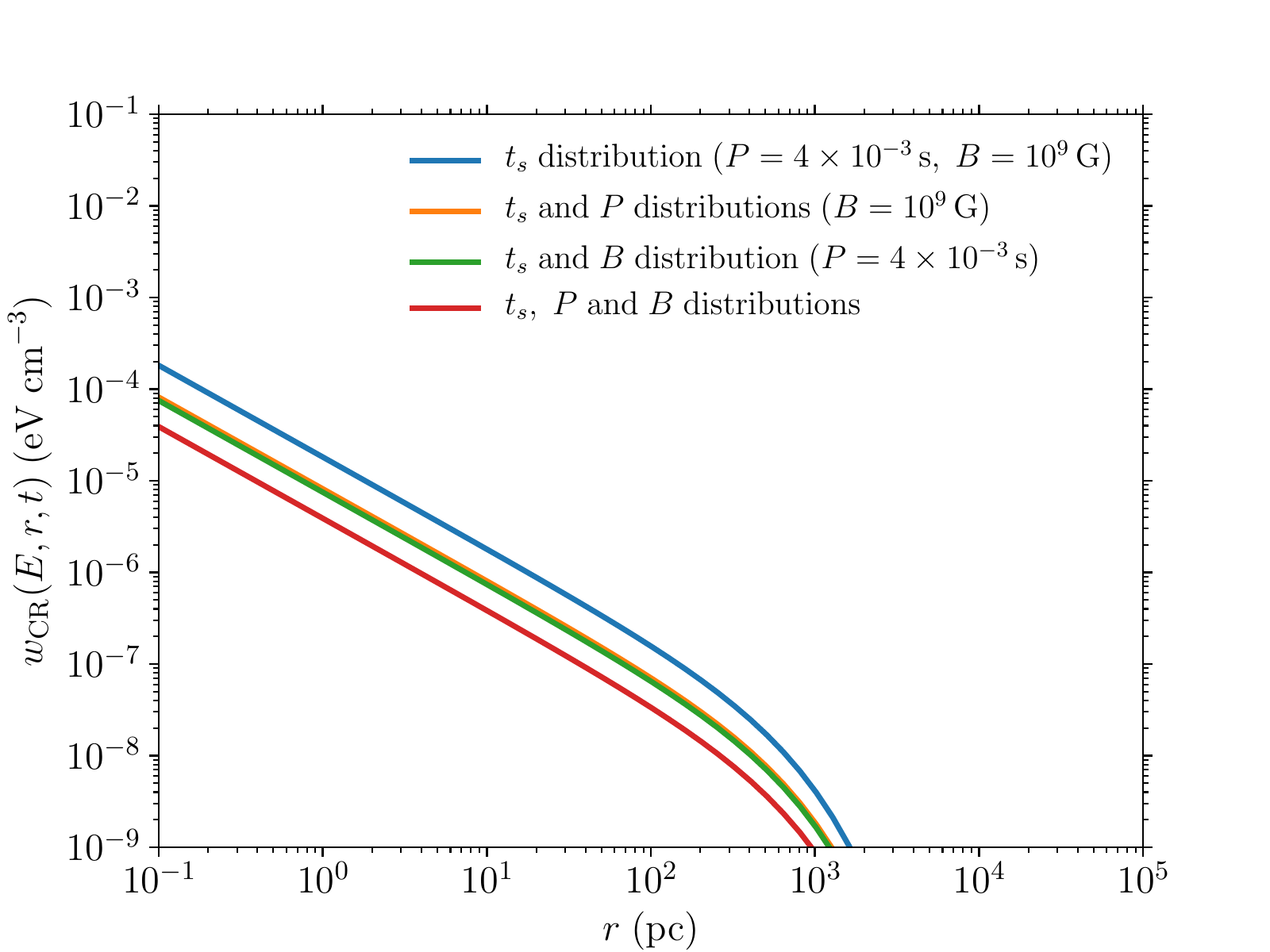}
\includegraphics[width=0.49\textwidth]{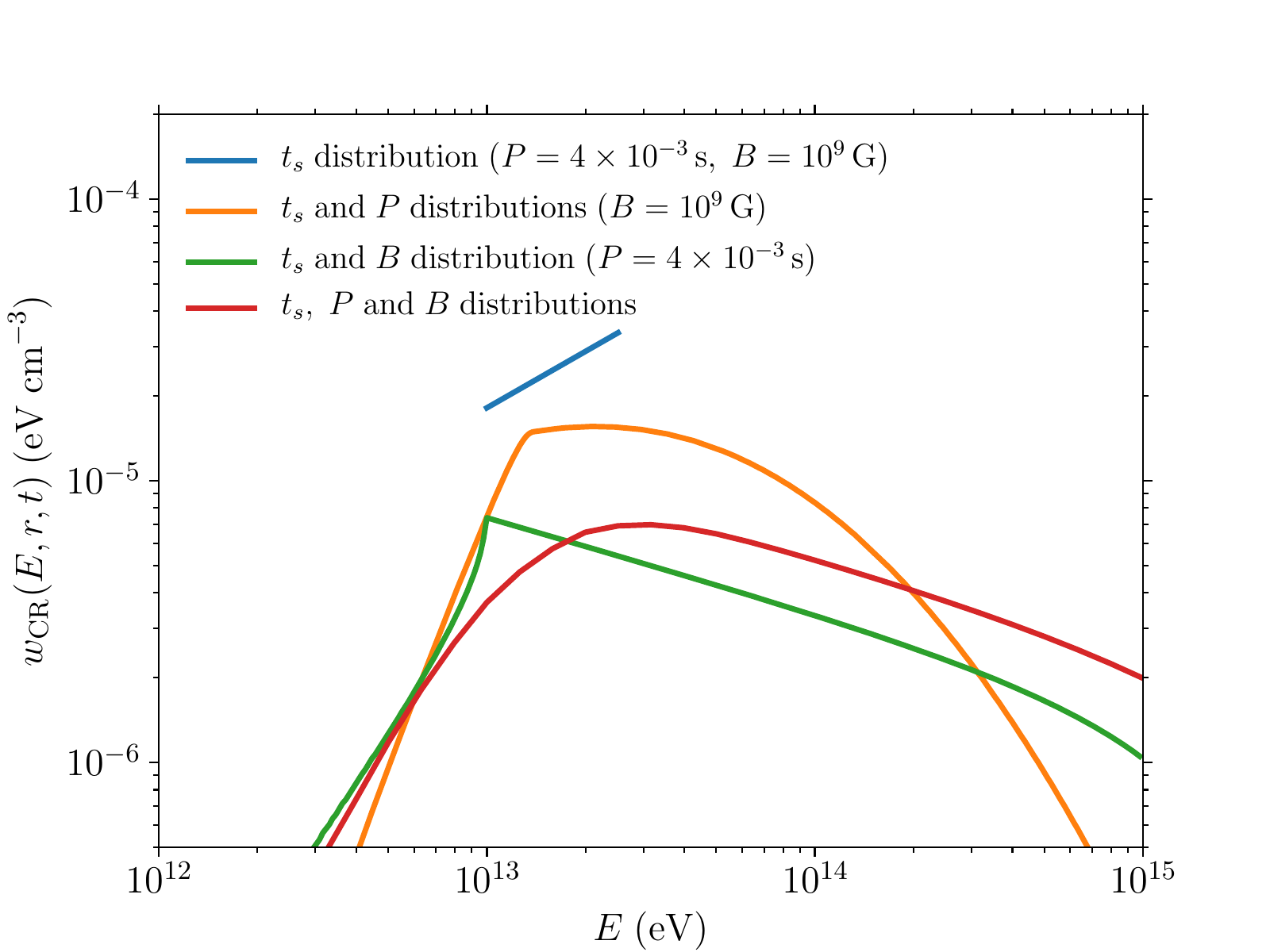}
\caption{Cosmic-ray density for $\eta_{\rm acc} = 1$ and $t=T_{s,{\rm max}}=10^{17}\,{\rm s}$, as a function of distance for $E=10^{13}\,{\rm eV}$ ({\it left}) and as a function of energy for $r=1\,{\rm pc}$ ({\it right}). We show the cosmic-ray densities integrated on: $t_s$ distribution, for $P=4\times10^{-3}\,{\rm s}$ and $B=10^9\,{\rm G}$ (blue); $t_s$ and $P$ distributions, for $B=10^9\,{\rm G}$ (orange); $t_s$ and $B$ distributions, for $P=4\times10^{-3}\,{\rm s}$ (green); $t_s$, $P$ and $B$ distributions (red).}\label{Fig:Int_PBt_dist}
\end{figure}

We study the influence of the initial spin and dipole magnetic field distributions on the cosmic-ray densities, for the uniform power-law injection model, illustrated in figure~\ref{Fig:Int_PBt_dist_const} and the transient monoenergetic injection model in figure~\ref{Fig:Int_PBt_dist}.

\section{Differential cross sections for the gamma-ray production}\label{App:dsigma}

\begin{figure}[ht]
\centering
\includegraphics[width=0.49\textwidth]{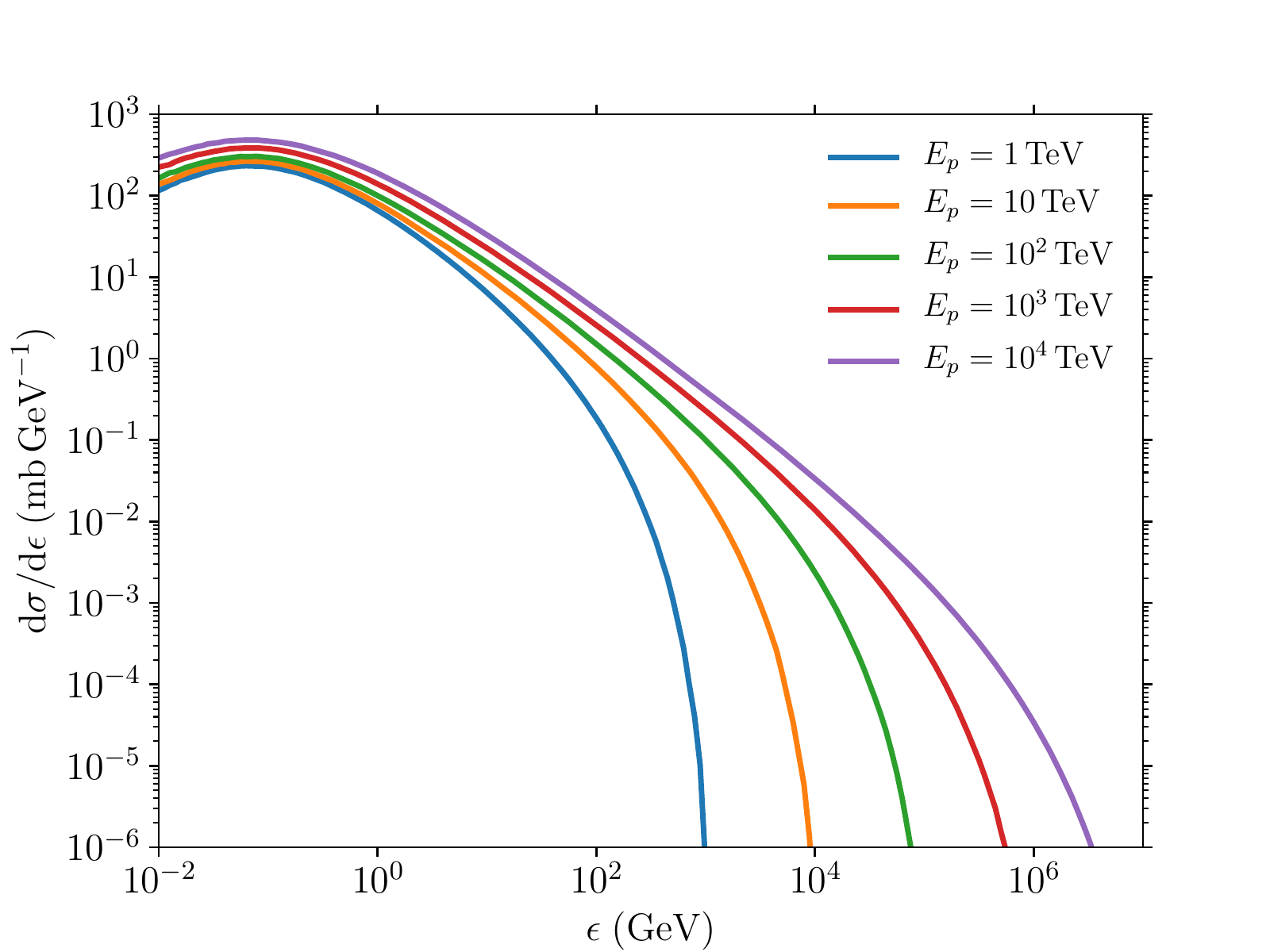}
\caption{Differential cross sections for the gamma-ray production  ${\rm d}\sigma_{pp,\gamma}(\epsilon,E)/{\rm d}\epsilon\,({\rm mb\,GeV}^{-1})$, as a function of gamma-ray energy $\epsilon\,({\rm GeV})$, for various proton energies $E=1-10^4\,{\rm TeV}$.}\label{Fig:dsigma}
\end{figure}

The differential cross sections for the gamma-ray production are illustrated in Fig.~\ref{Fig:dsigma}, for different proton energies between $E=1\,{\rm TeV}$ and $E=10^4\,{\rm TeV}$. We note that in the energy range of interest (above $\epsilon = 1\,{\rm TeV}$), the differential cross section shows a strong dependence on the incident proton energy since the maximum energy is directly linked to the latter. In \cite{Kelner:2006tc}, these distributions have been parametrized as a function of the fraction of energy $x=\epsilon/E$ to remove this explicit dependence. Since for our study we are not interested in the detailed contributions of each hadronic component like in \cite{Kelner:2006tc}, we preferred a more straightforward approach using the up-to-date hadronic interaction model EPOS LHC \cite{Werner06,Pierog:2013ria} now widely used to study soft QCD results at LHC and air showers. The differential cross section for the photon production is simulated directly taking into account the decay of all unstable particles (mainly neutral pions and eta resonances) at different energies and then interpolated for the calculation of the integral in eq.~\ref{eq:ngamma}.

\section{Influence of the diffusion coefficient}\label{App:D_impact}

\begin{figure}[ht]
\centering
\includegraphics[width=0.49\textwidth]{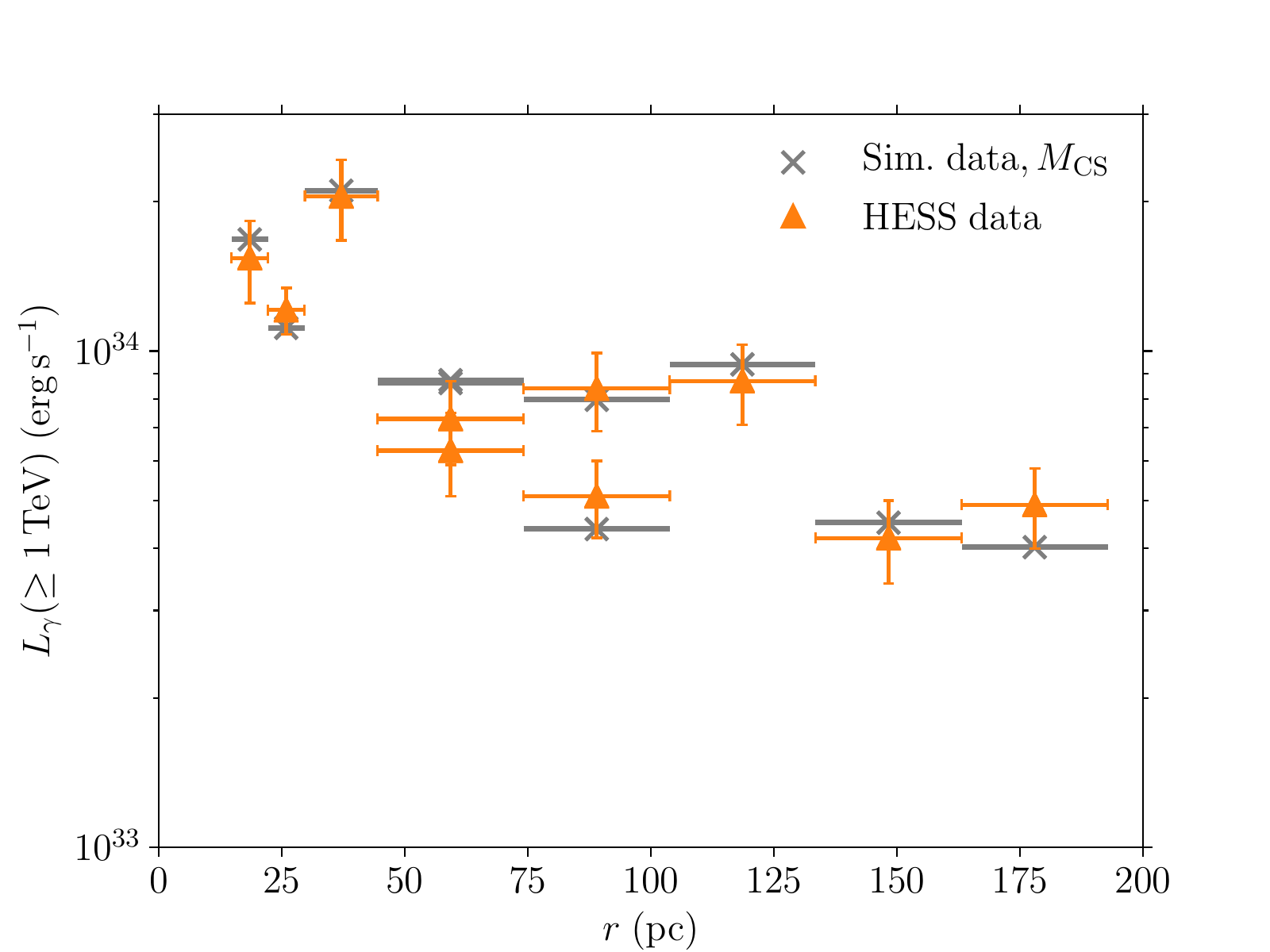}
\includegraphics[width=0.49\textwidth]{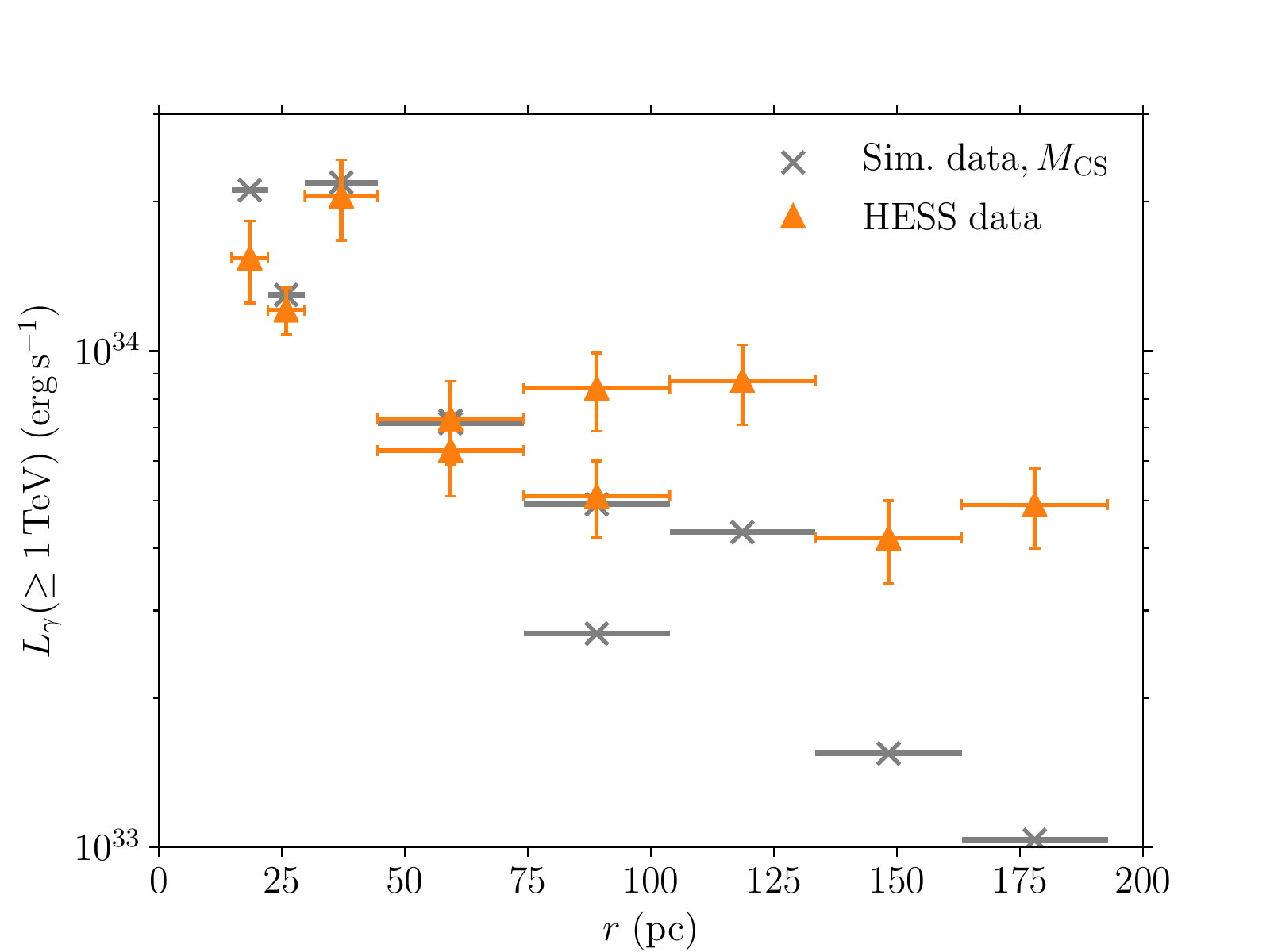}
\caption{Gamma-ray luminosity as a function of the distance to the galactic center, for our model (grey crosses) and measured by H.E.S.S. (orange triangles). We compare the results obtained for the diffusion coefficient $D(E)$ of Eq.~(\ref{eq:diff}) (left) and for $D(E)/100$ (right).}\label{Fig:D_comp}
\end{figure}

We present in figure~\ref{Fig:D_comp} the impact of the diffusion coefficient on the radial extent of the gamma-ray diffuse emission. For this purpose we compare two different diffusion coefficients, $D(E)$ and $D(E)/100$, where $D(E) = 10^{28} D_{28} \left(R/3\,{\rm GV}\right)^{\delta} \rm cm^{2}\, s^{-1} $, $R=E/Z$, $D_{28}/H_{\rm kpc}= 1.33$ and $\delta=1/3$ (see Eq.~(\ref{eq:diff})). Whereas the diffusion coefficient $D(E)$ allows to fit correctly the H.E.S.S. data, a decrease of the diffusion coefficient leads to higher gamma-ray luminosities at shorter distances and smaller gamma-ray luminosities at larger distances, and therefore does not allow to fit the H.E.S.S. data.

\bibliography{GCTeVbib}

\end{document}